\documentclass[lettersize,journal]{IEEEtran}
\usepackage{amsmath,amsfonts}
\usepackage{amsmath,amssymb,amsthm}
\usepackage{amssymb}
\usepackage{algorithm}
\usepackage{algpseudocode}
\usepackage{array}
\usepackage[caption=false,font=normalsize,labelfont=sf,textfont=sf]{subfig}
\usepackage{textcomp}
\usepackage{stfloats}
\usepackage{url}
\usepackage{verbatim}
\usepackage{graphicx}
\usepackage{cite}
\usepackage{makecell}  
\usepackage{tabularx,booktabs,makecell}
\usepackage{tikz}
\usetikzlibrary{arrows.meta, positioning, shapes, calc}

\usepackage{booktabs}
\usepackage{multirow}
\usepackage{adjustbox}
\usepackage{pifont}

\newcommand{\cmark}{\ding{51}}
\newcommand{\dash}{--}

\begin{document}

\title{Self-Interference-Aware AS-Assisted Tri-Hybrid Beamforming \\ for Full-Duplex Massive MIMO}
\author{Yuanzhe Gong\IEEEauthorrefmark{1},~\IEEEmembership{ Member,~IEEE}, Cheng-Jie Zhao\IEEEauthorrefmark{2},~\IEEEmembership{Student Member,~IEEE}, Tho Le-Ngoc\IEEEauthorrefmark{1},~
\IEEEmembership{Life~Fellow,~IEEE}

\thanks{Y. Gong, and T. Le-Ngoc are with the Department of Electrical and Computer Engineering, McGill University, Montreal, Canada. C.-J. Zhao is with the Department of Electrical and Computer Engineering, The University of Hong Kong, Hong Kong (Emails:
 yuanzhe.gong@mail.mcgill.ca, chengjie\_zhao@connect.hku.hk, tho.le-ngoc@mcgill.ca).
 }}
 
\markboth{Journal of \LaTeX\ Class Files,~Vol.~, No.~, May~2026}
{Shell \MakeLowercase{\textit{et al.}}: A Sample Article Using IEEEtran.cls for IEEE Journals}

\maketitle

\begin{abstract}
This paper proposes a tri-hybrid beamforming (tri-HBF) scheme with antenna-selection (AS)-based reconfigurable sub-arrays for full-duplex (FD) massive multiple-input multiple-output (mMIMO) systems. A sub-connected HBF architecture is adopted, where AS is performed in a group-wise manner to avoid excessive switch-network and routing complexity. 
An alternating optimization (AO) algorithm is developed to jointly optimize the i) active antenna subsets considering a self-interference (SI)-aware utility, ii) analog beamformers through projected gradient ascent (PGA), iii) digital precoders/combiners via SI-aware regularized zero-forcing (RZF) and minimum mean-square error (MMSE) updates, and iv) DL/UL power allocation by successive convex approximation (SCA). 
To capture realistic electromagnetic coupling in FD mMIMO operation, experimental SI channels based on an \(8\times8\) Tx-\(8\times8\) Rx FD array prototype are incorporated into the study. 
The proposed AS-aided tri-HBF optimization scheme exhibits robust convergence across various base station configurations and effectively balances desired-signal enhancement, SI mitigation, and multi-user interference suppression in FD mMIMO operation. 
Illustrative results show that selective activation can outperform full-array activation, achieving a \(21.3\%\) higher average sum-rate and a more consistent performance across user realizations, with power-efficiency benefits by reducing the active paths. 
A comprehensive study is conducted to characterize how the number of activated antennas affects the achievable rate, user-channel coherence, and SI suppression gain.
Compared with various selection baselines, it achieves a \(45.1\%\) improvement in average sum-rate, with average DL and UL rate gains of \(36.9\%\) and \(82.9\%\), respectively. In addition, beam-level isolation better than \(63\) dB is achieved, further confirming the effectiveness of the proposed SI-aware design.
\end{abstract}


\section{Introduction}
\IEEEPARstart{F}{ull}-duplex (FD) communication enables simultaneous transmission and reception over the same time-frequency resources, thereby providing the potential to nearly double the spectral efficiency of conventional half-duplex (HD) systems while reducing transmission latency \cite{sabharwal2014band, le2024full}.
In parallel, massive multiple-input multiple-output (mMIMO) systems employ large-scale antenna arrays to enhance spatial multiplexing capability, support simultaneous multi-user (MU) transmission, and achieve order-of-magnitude capacity gains with improved radiated energy efficiency \cite{marzetta2010noncooperative,rusek2012scaling}. 
The integration of FD and mMIMO is therefore a promising solution for meeting the ever-increasing demands for spectral efficiency, capacity, and energy-efficient connectivity in next-generation wireless networks \cite{nguyen2020spectral,le2026half}.
Meanwhile, a major challenge limiting practical deployment of FD mMIMO systems is self-interference (SI), caused by the leakage of the transmitter (Tx) signal into the co-located receiver (Rx). 
Since the Tx and Rx arrays are placed in close proximity at an FD base station (BS), the SI channel typically experiences much lower propagation loss than the desired user-BS channels \cite{zhang2016full}. As a result, the SI power can be orders of magnitude stronger than the desired signal, severely degrading Rx performance and potentially driving RF front-end components, such as the low-noise amplifier (LNA) and analog-to-digital converter (ADC), into nonlinear operation.

To achieve uplink (UL) performance comparable to HD operation, substantial SI mitigation is required.
For example, with a transmit power of \(P_T=30\,\mathrm{dBm}\), a noise power spectral density (PSD) of \(-174\,\mathrm{dBm/Hz}\), and a bandwidth of \(20\,\mathrm{MHz}\), approximately \(131\,\mathrm{dB}\) of SI suppression is required \cite{Report_5G_Macro_PL_Rel_17}.
Achieving such a high suppression level poses a stringent challenge, and multi-stage SI mitigation is typically employed. 
In particular, SI isolation aims to prevent the leaked SI wave from being captured by the local Rx through RF isolation structures \cite{farahani2010mutual, dadgarpour2016mutual, YG_ACCESS_2}, antenna design and spatial-domain suppression \cite{duarte2013design,debaillie2014analog, nawaz2017dual,YG_ACCESS_1}. The remaining residual SI is then further reduced by self-interference cancellation (SIC) techniques in the RF and/or digital domains, including digital-replica-based cancellation \cite{ahmed2015all,liu2023joint}, adaptive filter-weight optimization, and dedicated cancellation-circuit designs \cite{hong2022frequency}.

Recent advances in reconfigurable multi-antenna technologies, such as antenna selection (AS) \cite{wilson2017antenna,fidan2018performance,jang2016antenna,mahmood2024achieving, gong2026subarray,gong2026JointICCE}, movable antennas (MAs) \cite{ding2024movable,li2025joint,zhao2025joint}, fluid-antenna systems (FASs) \cite{hong2025fluid,skouroumounis2023fluid}, and reconfigurable intelligent surface (RIS)-assisted architectures \cite{wu2023ris,zhang2023ris,yang2023reconfigurable,tewes2022full,guo2023joint,le2023ris}, provide new degrees of freedom (DoFs) for FD MIMO and mMIMO design. By enabling adaptive antenna, aperture, or propagation-path configurations, these techniques can improve SI mitigation under user- and channel-dependent conditions, while maintaining the desired-link performance \cite{yang2015efficient,fidan2018performance,zhu2022antenna,wilson2017antenna,jang2016antenna,ding2024movable,gao2015massive,li2025joint,zhao2025joint,wu2023ris,zhang2023ris,yang2023reconfigurable,tewes2022full,guo2023joint,le2023ris,hong2025fluid,skouroumounis2023fluid}.
For instance, MA-enabled FD systems have been optimized by jointly designing antenna positions, beamformers, and power allocation using alternating optimization (AO) frameworks for secrecy-rate maximization \cite{ding2024movable}, and binary particle swarm optimization (BPSO) combined with successive convex approximation (SCA) and difference-of-convex (DoC) programming for FD integrated sensing and communication scenarios (ISAC) scenarios \cite{li2025joint}. These studies show that antenna-position reconfiguration provides additional spatial DoFs for shaping desired and interference channels.
Additionally, in \cite{hong2025fluid}, an FAS-assisted SIC framework was proposed, where an Rx-side FAS selects the port with the minimum residual SI by exploiting the spatial variation between the desired forward channel and the loopback SI channel. 
In \cite{skouroumounis2023fluid}, large-scale FAS-aided FD cellular networks were developed with strongest-port selection, sequential LMMSE channel estimation, and UL fractional power control under spatially correlated FA ports and residual loop interference, achieving an average sum-rate improvement of 45\% over static FD systems.
Moreover, RIS-assisted FD systems exploit propagation-domain reconfigurability for SI mitigation and desired-channel enhancement by tuning the electromagnetic (EM) response of low-cost reflecting elements \cite{wu2019towards}. 
Existing studies have investigated RIS-assisted FD systems from several perspectives, including joint BS/user precoding and RIS phase-shift optimization for multi-cell FD sum-rate maximization under SI and inter-cell interference~\cite{chen2023next}, experimental RIS-assisted analog-domain SIC at the BS with an additional 59 dB SI suppression using a 256-element prototype~\cite{tewes2022full}, and joint RIS phase and SI-mitigation precoder design with performance bounds for RIS-assisted FD transceivers~\cite{zhang2023ris}.

Despite their potential, MA- and FAS-based FD designs face practical scalability challenges. MAs require physical or electronically controlled antenna displacement, whereas FASs rely on specialized fluid, liquid-metal, or switchable-port structures. 
These mechanisms introduce additional control and switching overhead, calibration complexity, finite reconfiguration speed, and hardware-reliability concerns. 
When extended to BS-side large-scale MIMO, the overhead in feeding networks, control circuitry, switching, and CSI acquisition grows with the number of movable elements or candidate ports.
Moreover, wavelength-scale movement or port-switching regions can hinder compact array integration and limit RF-domain spatial multiplexing flexibility, making their deployment in contemporary BS infrastructures challenging.
On the other hand, RIS-assisted designs offer propagation-domain reconfigurability at the cost of additional external hardware and increased joint-optimization complexity. 
Their implementation requires accurate cascaded BS-RIS-user/SI channel acquisition, reliable and accessible RIS control and placement, and BS-RIS synchronization. 
The reflected decoupling signal is highly sensitive to propagation distance and reflection geometry, making it difficult to maintain a strength comparable to that of the dominant direct Tx-Rx SI path. 

Since antenna elements do not contribute equally in practical propagation environments, AS provides an effective means to reconfigure the EM aperture by activating only the most informative and well-conditioned antennas \cite{gao2015massive,li2026tri}. 
Several studies have investigated AS for FD systems \cite{yang2015efficient,wilson2017antenna,jang2016antenna,fidan2018performance,zhu2022antenna,mahmood2024achieving,gong2026subarray,gong2026JointICCE}. 
A joint relay and antenna-mode selection scheme was proposed for FD relay networks in \cite{yang2015efficient}, where each relay adaptively selects its Tx/Rx antennas based on channel conditions and signal-interference-plus noise ratio (SINR), thereby suppressing the SI-induced error floor and providing additional spatial diversity. 
For FD MIMO systems, low-complexity AS algorithms based on channel magnitude, orthogonality, determinant, and greedy selection criteria were developed in \cite{jang2016antenna,wilson2017antenna}, achieving near-optimal performance and up to 15\% gain over conventional schemes \cite{jang2016antenna}.
However, existing FD AS studies \cite{yang2015efficient, wilson2017antenna, jang2016antenna, fidan2018performance,zhu2022antenna} mainly focus on small-scale MIMO, relay, or single-antenna-per-user scenarios with theoretical or simulation-based channel models, and thus do not fully capture the coupled multi-user interference (MUI)-SI effects or exploit the spatial diversity of FD mMIMO arrays.

For large-scale mMIMO systems, AS can exploit spatial non-uniformity across a large number of candidate antennas to reduce RF-chain usage and system complexity while maintaining competitive performance. 
Although mainly focused on HD scenarios, received-SNR-based simulation results have shown that AS can substantially outperform non-selection schemes \cite{kermoal2002stochastic,gkizeli2014maximum}.
Measurement results using 128-element linear and cylindrical arrays further confirmed significant element-wise power variations in propagation environments, where simple received-power-based AS can approach convex-optimization benchmarks \cite{gao2015massive}.
Furthermore, recent tri-hybrid mMIMO architectures expand the available spatial DoFs by distributing spatial processing across digital beamforming, RF beamforming, and antenna-level EM reconfigurability \cite{heath2025tri,li2026tri}. 
By alternatingly optimizing the additional EM-domain precoder with conventional RF/baseband (BB) beamforming, these architectures provide a promising path toward energy-efficient large-aperture mMIMO with improved spectral/energy efficiency and reduced hardware complexity \cite{li2026tri}.
However, existing large-scale AS and tri-hybrid/reconfigurable mMIMO studies \cite{kermoal2002stochastic,gkizeli2014maximum,gao2015massive,heath2025tri,li2026tri} mainly focus on HD scenarios and \cite{kermoal2002stochastic,gkizeli2014maximum,heath2025tri,li2026tri} rely on analytical or simulation-based channels, leaving selection-based FD designs insufficiently validated under experimentally measured SI channels. In FD arrays, the close Tx-Rx spacing induces strong and rapid spatial variation in the SI channel, which creates selection diversity for reducing Tx-Rx coupling while preserving desired-link quality.
Although FD mMIMO HBF with sub-array selection (SAS) and measured SI channels has been studied in~\cite{mahmood2024achieving,gong2026subarray}, these works mainly focus only on RF-beamformer optimization with fixed sub-array configurations for SI and MUI mitigation. The adopted particle swarm optimization (PSO) method is suitable when the number of candidate sub-arrays is limited; however, it becomes difficult to scale to element-level AS in large-scale arrays due to the significantly enlarged mixed discrete-continuous search space.
Moreover, fixed sub-array-level selection may lead to inefficient antenna utilization and potential RF-chain rerouting, as many candidate sub-arrays are required for selection diversity but only a very limited subset is activated.
Meanwhile, although a PSO-based AS scheme was developed in~\cite{gong2026JointICCE}, it relies on a fully digital beamformer with an RF-chain-to-all-antenna switching network. This architecture provides flexibility but removes the RF-domain beamforming gain available in HBF and requires excessive switching/routing complexity, making it less suitable for scalable large-array implementation.
Table~\ref{tab:literature_comparison} summarizes and compares related reconfigurable architectures.

\begin{table*}[t]
\centering
\caption{Comparison of Reconfigurable-Technologies-Enabled FD/MIMO Systems}
\label{tab:literature_comparison}
\scriptsize
\setlength{\tabcolsep}{2.0pt}
\renewcommand{\arraystretch}{0.8}
\begin{adjustbox}{width=\textwidth}
\begin{tabular}{c |c| c c c c | c c c c c | >{\centering\arraybackslash}p{2.85cm}|
>{\centering\arraybackslash}p{3.95cm}}
\toprule
\multirow{2}{*}{Ref.}
& \multirow{2}{*}{Mode}
& \multicolumn{4}{c}{Reconfigurable Technology}
& \multicolumn{5}{c}{Design Components}
& \multirow{2}{*}{{Channel /Validation}}
& \multirow{2}{*}{\makecell[c]{Main Method\\of Optimization}} \\
\cmidrule(lr){3-6} \cmidrule(lr){7-11}
& & AS & MA & FAS & RIS
& MIMO & Large Array
& \makecell[c]{RF BF}
& \makecell[c]{Digital BF}
& Power Allocation
& & \\
\midrule

\cite{wilson2017antenna}
& FD & \cmark & \dash & \dash & \dash
& \cmark & \dash & \dash
& \dash
& \dash
& Simul. (Rayleigh)
& Magnitude/null/determinant AS \\ \midrule

\cite{fidan2018performance}
& FD & \cmark & \dash & \dash & \dash
& \cmark & \dash & \dash & \dash & \cmark
& Simul. (Rayleigh)
&Sum outage probability-based AS \\ \midrule

\cite{jang2016antenna}
& FD & \cmark & \dash & \dash & \dash
& \cmark & \dash & \dash & \dash & \dash
& Simul. (Rayleigh/Rician) 
& Exhaustive and greedy AS \\ \midrule

\cite{ding2024movable}
& FD & \dash & \cmark & \dash & \dash
& \cmark & \dash & \dash & \cmark & \cmark
& \dash
& AO (multi-velocity PSO, SCA) \\ \midrule

\cite{li2025joint}
& FD & \dash & \cmark & \dash & \dash
& \cmark & \dash & \dash & \cmark & \cmark
& \dash
& BPSO, DoC, SCA \\ \midrule

\cite{hong2025fluid}
& FD & \makecell[c]{Port sel.} & \dash & \cmark & \dash
& \dash & \dash & \dash & \dash & \dash
& Simulations
& Residual SI bound and port sel. \\ \midrule

\cite{skouroumounis2023fluid}
& FD & \makecell[c]{Port sel.} & \dash & \cmark & \dash
& \dash & \dash & \dash & \dash & \cmark
& Stochastic-geometry 
& Stochastic geometry, LMMSE  \\ \midrule 

\cite{zhang2023ris}
& FD & \dash & \dash & \dash & \cmark
& \cmark & \dash 
& \dash
& \cmark 
& \cmark
& Simulations
& AO, Riemannian conjugate gradient \\ \midrule 

\cite{tewes2022full}
& FD & \dash & \dash & \dash & \cmark
& \dash & \dash 
& \dash 
& \dash & \dash
& RIS/FD radio testbed
& Greedy binary RIS search \\ \midrule

\cite{le2023ris}
& FD & \dash & \dash & \dash & \cmark
& \cmark & \dash & \dash
& \cmark
& \cmark
& Simulations
& Inner approx. block coordinate ascent\\ \midrule 

\cite{gao2015massive}
& HD & \cmark & \dash & \dash & \dash
& \cmark & \makecell[c]{128-ant.}
& \cmark
& \cmark 
& \cmark 
& Measured channels 
& Convex/power-based AS \\ \midrule

\cite{li2026tri}
& HD & \cmark & \dash & \dash & \dash
& \cmark &  \cmark 
& \cmark
& \cmark
& \dash
& Simul. (mmWave)
& Tri-loop AO  \\ \midrule  

\cite{shan2026fluid}
& \makecell[c]{FD/HD}
& \makecell[c]{Port sel.} & \dash & \cmark & \dash
& \dash & \dash & \dash & \dash & \cmark
& Simul. (Rayleigh/Jakes)
& AO, ternary search \\ \midrule 

\cite{tang2025full}
& FD & \dash & \dash & \cmark & \dash
& \cmark & \dash & \dash & \cmark & \cmark
& Simul. (Multipath)
& AO (fractional programming, SCA) \\ \midrule 

\cite{abdullah2022low}
& HD & \cmark & \dash & \dash & \cmark
& \cmark & \cmark & \dash & \cmark & \dash
& Simul. (Rayleigh/Rician)
& Discrete RIS phase design \\ \midrule

\cite{gao2017massive}
& HD & \cmark & \dash & \dash & \dash
& \cmark & \cmark & \dash & \dash & \dash
& Simul. (Rayleigh)
& Greedy/branch-bound AS  \\ \midrule

\textbf{This work}
& \textbf{FD}
& \cmark
& \dash & \dash & \dash
& \cmark
& \makecell[c]{64Tx/64Rx}
& \cmark
& \cmark
& \cmark
& \textbf{Measured SI channel}
& \textbf{AO (PGA,RZF,MMSE,SCA)} \\ 
\bottomrule
\end{tabular}
\end{adjustbox}
\end{table*}

\subsection{Contributions and Organization}
Motivated by the above observations, this paper proposes an AO-based framework for AS-assisted tri-hybrid beamforming (tri-HBF) FD mMIMO systems. To the best of our knowledge, this is the first study to incorporate experimentally measured SI channels while jointly optimizing AS across large-scale arrays, hybrid beamforming (HBF), and power allocation for FD mMIMO. The main contributions are summarized as follows:
\begin{itemize}
   \item \textbf{AS-aided tri-HBF design with a practical sub-connected architecture:} An AO algorithm is developed to jointly optimize i) the active antenna subsets using a customized SI-aware utility, ii) the analog beamformers via projected gradient ascent (PGA), iii) the digital precoders/combiners through SI-aware regularized zero-forcing (RZF) and minimum mean-square error (MMSE) updates, and iv) the DL/UL power allocation using SCA. To improve hardware practicality and complexity, a sub-connected HBF architecture is adopted. Thus, AS is performed in a group-wise manner to avoid excessive switch-network and RF-routing complexity while reducing the number of RF chains required to interface the analog RF stage with the digital BB. The proposed AO-based scheme demonstrates fast and robust convergence across different activated-antenna and RF-chain configurations. 

    \item \textbf{Evaluation with measurement-based SI channels:}
To account for practical EM coupling and SI effects in FD mMIMO operation, an experimentally measured SI channel obtained from a large-scale \(8\times8\) Tx- \(8\times8\) Rx FD antenna-array prototype is incorporated into the evaluation. Compared with various AS baselines, the proposed scheme achieves a \(45.1\%\) improvement in the average sum-rate, with average DL and UL rate gains of \(36.9\%\) and \(82.9\%\), respectively. The proposed design provides beam-level isolation exceeding \(63\) dB, corresponding to more than \(10\) dB isolation improvement over both the desired-signal-only and random selection schemes.

  \item \textbf{Comprehensive empirical characterization of selective antenna activation across large-scale arrays:}
A comprehensive study is conducted to characterize the impact of the number of activated antennas on the achievable FD sum rate, user-channel coherence, and effective SI channel mitigation performance. The results show that selective antenna activation can outperform full-array activation, achieving a \(21.3\%\) higher average sum-rate and a more uniform performance distribution across user realizations. Moreover, selective activation offers potential power-efficiency benefits by reducing the number of active RF paths. The effective beam-level SI suppression alleviates receiver dynamic-range limitations and RF-chain distortion, confirming that the proposed tri-HBF design effectively balances desired-signal enhancement and SI suppression in FD mMIMO.
\end{itemize}

The rest of this paper is organized as follows. Section~\ref{sec:system_model} presents the system model of the proposed tri-HBF architecture with reconfigurable sub-arrays for FD mMIMO systems. Section~\ref{sec:methodology} introduces the proposed AO-Based Tri-HBF optimization framework. Section~\ref{Illustrative} presents illustrative results based on the experimentally measured large-scale-array SI channel and the theoretical BS-user multipath channel, demonstrating the effectiveness of the design and characterizing the impact of the number of activated antenna elements on FD mMIMO performance. Finally, Section~\ref{Conclusions} concludes the paper.

\section{AS-Assisted Tri-HBF for FD mMIMO Systems}\label{sec:system_model}
\begin{figure}[!t]
\centerline{\includegraphics[width=0.7\linewidth]{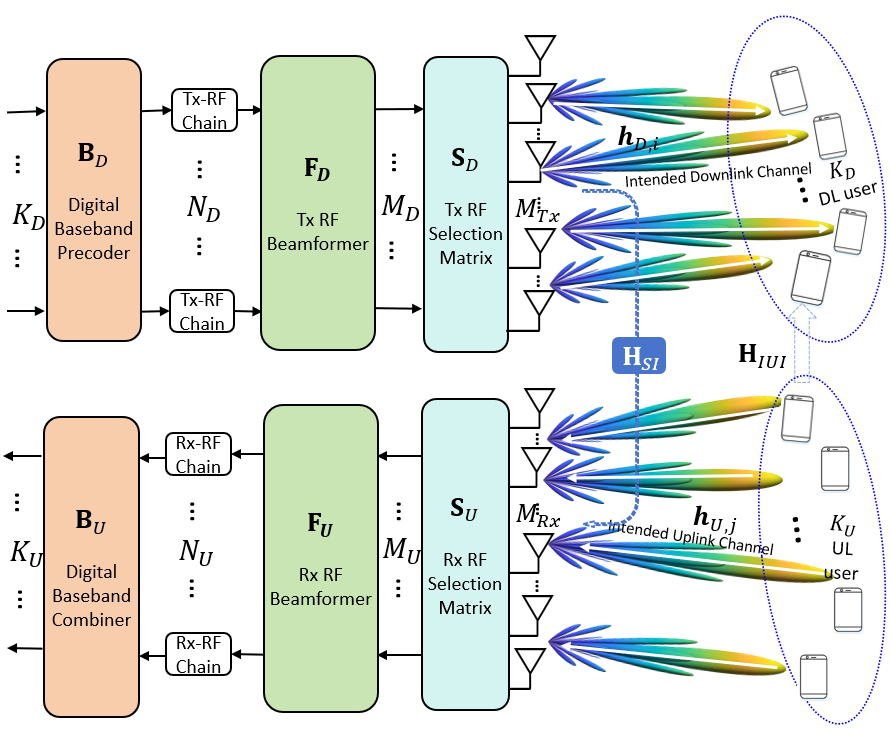}}
\vspace{-0.3cm}
\caption{Tri-HBF with reconfigurable sub-arrays for FD mMIMO.}
\vspace{-0.3cm}
\label{AO_HBF}
\end{figure}

As illustrated in Fig.~\ref{AO_HBF}, an FD MU-mMIMO BS employing joint AS-enabled reconfigurable sub-arrays, HBF, and power allocation is considered.
The BS is equipped with a co-located large-scale array comprising $M_{\mathrm{Tx}}$ Tx and $M_{\mathrm{Rx}}$ Rx antennas.
An RF-switch-based AS stage activates \(M_D\) Tx and \(M_U\) Rx elements to simultaneously serve \(K_D\) single-antenna DL users and \(K_U\) single-antenna UL users over the same time-frequency resource.
On the DL side, the BS employs $N_D$ RF chains satisfying $K_D \le N_D \le M_D \le M_{\mathrm{Tx}}$, and adopts an analog RF precoder $\mathbf F_D \in \mathbb C^{M_D \times N_D}$ followed by a digital BB precoder $\mathbf B_D=[\mathbf b_{D,1},\dots,\mathbf b_{D,K_D}] \in \mathbb C^{N_D \times K_D}$.
Similarly, on the UL side, the BS employs $N_U$ RF chains satisfying $K_U \le N_U \le M_U \le M_{\mathrm{Rx}}$, and adopts an analog RF combiner $\mathbf F_U \in \mathbb C^{M_U \times N_U}$ followed by a digital BB combiner $\mathbf B_U=[\mathbf b_{U,1},\dots,\mathbf b_{U,K_U}] \in \mathbb C^{N_U \times K_U}$.
By employing a reduced number of RF chains, the HBF significantly reduces RF-chain cost, power consumption, and overall hardware complexity, while preserving spatial processing capability.

\subsection{AS-Enabled Pre-Connected Reconfigurable Sub-array}
Let $\mathcal{Q}_D \subset \{1,\dots,M_{\mathrm{Tx}}\}$ and $\mathcal{Q}_U \subset \{1,\dots,M_{\mathrm{Rx}}\}$ denote the selected Tx and Rx antenna index sets, respectively, where $|\mathcal{Q}_D|=M_D$ and $|\mathcal{Q}_U|=M_U$.
A fully flexible antenna-to-RF-chain connection would require an excessively large-scale switch network and long RF routing paths, leading to high insertion loss, calibration complexity, control overhead, and hardware cost. 
To improve implementation practicality, a pre-connected sub-array architecture is adopted, where each RF chain is connected only to a predefined group of antenna elements, and AS is performed locally within each sub-array group.
Accordingly, the Tx and Rx arrays are partitioned into \(N_D\) and \(N_U\) disjoint sub-arrays, respectively. 
Under the equal-size grouping assumption, the Tx and Rx sub-array sizes are $M_{D,s}=\frac{M_{\mathrm{Tx}}}{N_D}$, and $M_{U,s}=\frac{M_{\mathrm{Rx}}}{N_U}$.
Let \(\{\mathcal{G}_{D,n}\}_{n=1}^{N_D}\) and \(\{\mathcal{G}_{U,n}\}_{n=1}^{N_U}\) denote the predefined Tx and Rx sub-array index sets. 
For given \(M_D\) and \(M_U\), exactly $L_D=\frac{M_D}{N_D}$, and $L_U=\frac{M_U}{N_U}$ antennas are selected from each Tx and Rx sub-array, respectively. 
Thus, the selected antenna sets satisfy
\begin{subequations}
\begin{align}
|\mathcal{Q}_D \cap \mathcal{G}_{D,n}|=L_D,\quad n=1,\ldots,N_D,\\
|\mathcal{Q}_U \cap \mathcal{G}_{U,n}|=L_U,\quad n=1,\ldots,N_U.
\end{align}
\end{subequations}
The group-wise AS structure preserves the sub-connected HBF architecture while avoiding the routing and switching burden of a fully flexible selection network.

The selection matrices 
\(\mathbf S_D\in\{0,1\}^{M_{\mathrm{Tx}}\times M_D}\) and
\(\mathbf S_U\in\{0,1\}^{M_{\mathrm{Rx}}\times M_U}\) are constructed by collecting the columns of
\(\mathbf I_{M_{\mathrm{Tx}}}\) and \(\mathbf I_{M_{\mathrm{Rx}}}\) indexed by
\(\mathcal{Q}_D\) and \(\mathcal{Q}_U\), respectively. 
With group-wise ordering of the selected antennas, the DL RF precoder and UL RF combiner have block-diagonal structures
\begin{subequations}
\begin{align}
\mathbf F_D \hspace{-0.3em}&= \hspace{-0.3em} \mathrm{blkdiag}(\mathbf f_{D,1},\hspace{-0.1em}\dots,\mathbf f_{D,N_D}), \hspace{-0.4em} \text{ where } |[\mathbf f_{D,n}]_{\ell}|\hspace{-0.3em} = \hspace{-0.3em}\frac{1}{\sqrt{L_D}},\\
\mathbf F_U \hspace{-0.3em} &= \hspace{-0.3em} \mathrm{blkdiag}(\mathbf f_{U,1},\dots,\mathbf f_{U,N_U}), \text{ where } |[\mathbf f_{U,n}]_{\ell}|\hspace{-0.3em} = \hspace{-0.3em}\frac{1}{\sqrt{L_U}}.
\end{align}\label{RF_BF}
\end{subequations}
\hspace{-0.2em}$\mathbf f_{D,n}\in\mathbb C^{L_D\times 1}$ and
\(\mathbf f_{U,n}\in\mathbb C^{L_U\times 1}\) denote the RF beamforming vectors associated with the \(n\)th selected Tx and Rx sub-arrays, respectively. 
Since the RF beamformers are implemented using phase shifters, their nonzero entries satisfy the constant-modulus constraints.
Let \(\mathbf h_{D,i}^{(\mathrm{full})}\in\mathbb C^{M_{\mathrm{Tx}}\times 1}\) denote the DL channel from the full Tx array to DL user \(i\), and let 
\(\mathbf H_U^{(\mathrm{full})}\in\mathbb C^{M_{\mathrm{Rx}}\times K_U}\) collect the UL channels from the \(K_U\) UL users to the full Rx array. 
The full SI channel between the co-located Tx and Rx arrays is denoted by
\(\mathbf H_{\mathrm{SI}}^{(\mathrm{full})}\in\mathbb C^{M_{\mathrm{Rx}}\times M_{\mathrm{Tx}}}\). 
The selected antenna-domain channels are then given by
\begin{subequations}
\begin{align}
\bar{\mathbf H}_D
&=
[\bar{\mathbf h}_{D,1},\ldots,\bar{\mathbf h}_{D,K_D}]
\in\mathbb C^{M_D\times K_D},\\
\bar{\mathbf h}_{D,i}
&=
\mathbf S_D^T\mathbf h_{D,i}^{(\mathrm{full})}
\in\mathbb C^{M_D\times 1},\\
\bar{\mathbf H}_U
&=
\mathbf S_U^T\mathbf H_U^{(\mathrm{full})}
\in\mathbb C^{M_U\times K_U},\\
\bar{\mathbf H}_{\mathrm{SI}}
&=
\mathbf S_U^T\mathbf H_{\mathrm{SI}}^{(\mathrm{full})}\mathbf S_D
\in\mathbb C^{M_U\times M_D}.
\end{align}
\label{eq:effective_channels_AS}
\end{subequations}

\vspace{-2em}
\subsection{DL/UL-User Received Signals}
Let \(\mathbf P_D=\mathrm{diag}(p_{D,1},\ldots,p_{D,K_D})\) and \(\mathbf P_U=\mathrm{diag}(p_{U,1},\ldots,p_{U,K_U})\) denote the nonnegative DL and UL stream-power allocation matrices, respectively. 
The DL and UL data symbol vectors satisfy
\(\mathbf d_D\sim\mathcal{CN}(\mathbf 0,\mathbf I_{K_D})\) and
\(\mathbf d_U\sim\mathcal{CN}(\mathbf 0,\mathbf I_{K_U})\). 
Under the proposed tri-HBF scheme, the BS DL transmit signal is given by
\begin{equation}
\begin{aligned}
\mathbf x_D
&=
\mathbf S_D\mathbf F_D\mathbf B_D\mathbf P_D^{1/2}\mathbf d_D,\\
\mathbb E\!\left[\|\mathbf x_D\|^2\right]
&=
\mathrm{tr}\!\left(
\mathbf F_D\mathbf B_D\mathbf P_D\mathbf B_D^H\mathbf F_D^H
\right)
\leq P_D^{\mathrm{tot}},
\end{aligned}
\label{eq:xD_hbf}
\end{equation}
where the equality follows from \(\mathbf S_D^T\mathbf S_D=\mathbf I_{M_D}\).
Let \(\mathbf w_D\sim\mathcal{CN}(\mathbf 0,\sigma_D^2\mathbf I_{K_D})\) denote the additive DL noise vector. The stacked DL received signal is
\begin{equation}
\mathbf r_D=\bar{\mathbf H}_D^H\mathbf F_D\mathbf B_D\mathbf P_D^{1/2}\mathbf d_D
+\mathbf G\mathbf P_U^{1/2}\mathbf d_U
+\mathbf w_D,
\label{eq:rD_hbf}
\end{equation}
where \(\mathbf G\in\mathbb C^{K_D\times K_U}\) denotes the inter-user interference (IUI) channel from the UL users to the DL users, with \([\mathbf G]_{i,j}=g_{i,j}\). Accordingly, the received signal at DL user \(i\) is
\begin{equation}
\begin{aligned}
r_{D,i}
=&
\underbrace{
\bar{\mathbf h}_{D,i}^H\mathbf F_D\mathbf b_{D,i}
\sqrt{p_{D,i}}d_{D,i}
}_{\text{desired signal}}+
\underbrace{
\mathbf g_i^T\mathbf P_U^{1/2}\mathbf d_U
}_{\text{IUI}}
\\
&+
\underbrace{
\sum_{\tilde i\neq i}
\bar{\mathbf h}_{D,i}^H\mathbf F_D\mathbf b_{D,\tilde i}
\sqrt{p_{D,\tilde i}}d_{D,\tilde i}
}_{\text{DL MUI}}
+
w_{D,i},
\end{aligned}
\label{eq:rDi_hbf}
\end{equation}
where \(\mathbf g_i=[g_{i,1},\ldots,g_{i,K_U}]^T\in\mathbb C^{K_U\times 1}\) collects the IUI channels from all UL users to DL user \(i\).

Let \(\tilde{\mathbf w}_U\sim\mathcal{CN}(\mathbf 0,\sigma_U^2\mathbf I_{M_U})\) denote the Rx thermal noise. After RF and BB combining, the detected UL signal vector and the \(j\)-th UL scalar detected signal are
\begin{equation}
\begin{aligned}
\mathbf r_U
=&
\underbrace{
\mathbf B_U^H\mathbf F_U^H\bar{\mathbf H}_U
\mathbf P_U^{1/2}\mathbf d_U
}_{\text{desired signal + UL MUI}}
+
\underbrace{
\mathbf B_U^H\mathbf F_U^H\bar{\mathbf H}_{\mathrm{SI}}
\mathbf F_D\mathbf B_D\mathbf P_D^{1/2}\mathbf d_D
}_{\text{SI}}
\\
&+
\underbrace{
\mathbf B_U^H\mathbf F_U^H\tilde{\mathbf w}_U
}_{\text{post-combining noise}} .
\end{aligned}
\label{eq:rU_hbf}
\end{equation}
\begin{equation}
r_{U,j}
\hspace{-0.3em}=\hspace{-0.3em}
\mathbf b_{U,j}^H\mathbf F_U^H
\hspace{-0.1em}(
\bar{\mathbf H}_U\mathbf P_U^{1/2}\mathbf d_U
\hspace{-0.2em}+\hspace{-0.2em}
\bar{\mathbf H}_{\mathrm{SI}}\mathbf F_D\mathbf B_D
\mathbf P_D^{1/2}\mathbf d_D
\hspace{-0.2em}+\hspace{-0.2em}
\tilde{\mathbf w}_U).
\label{eq:rUj_hbf}
\end{equation}

The considered FD mMIMO system involves two different channel classes, which are modeled differently to balance tractability and realism in the study.

First, the intended BS-user channels are generated using a conventional far-field narrowband geometric multipath model. Specifically, the channel between the BS array and user \(i\) on link \(q\in\{D,U\}\) is modeled as a Rician channel:
\begin{equation}
\begin{aligned}
\mathbf h_{q,i}^{(\mathrm{full})}
\hspace{-0.2em}=\hspace{-0.2em}
\sqrt{M_q\beta_i}
&\bigg(
\Omega_i
\sqrt{\frac{K_R}{K_R+1}}
\alpha_{i,0}\mathbf a(\theta_{i,0},\psi_{i,0})
\\
&+
\sqrt{\frac{1}{\Omega_iK_R+1}}
\sum_{\ell=1+\Omega_i}^{L_i}
z_{i\ell}\mathbf a(\theta_{i\ell},\psi_{i\ell})
\bigg),
\end{aligned}
\label{eq:intended_channel}
\end{equation}
where \(M_q\in\{M_{\rm Tx},M_{\rm Rx}\}\) denotes the number of BS antenna elements for the corresponding Tx/Rx array, \(\Omega_i\in\{0,1\}\) is the line-of-sight (LoS) indicator, and \(\Omega_i=1\) and \(0\) correspond to LoS and non-line-of-sight (NLoS) realizations, respectively. The LoS indicator is generated according to a prescribed LoS probability, i.e., \(\Omega_i\sim\mathrm{Bernoulli}(P_{\rm LoS})\). Moreover, \(K_R\) is the Rician factor, \(\alpha_{i,0}\) is the unit-modulus LoS path coefficient, and \(\mathbf a(\theta,\psi)\) denotes the unit-norm array steering vector. The factor \(\sqrt{M_q}\) ensures that the average per-antenna channel power follows the large-scale path gain \(\beta_i\). The large-scale path gain is given by $\beta_i= \left(\frac{\lambda}{4\pi}\right)^2d_i^{-\eta}$,
\(\lambda\) is the wavelength, \(d_i\) is the BS-UE distance, and \(\eta\) is the path-loss exponent. The scattering coefficients satisfy $z_{i\ell}\sim \mathcal{CN}\left(0,\frac{1}{L_i-\Omega_i}\right)$.
When \(\Omega_i=1\), the channel contains one LoS component and \(L_i-1\) angularly spread NLoS components. When \(\Omega_i=0\), the LoS term vanishes and the channel is generated from \(L_i\) scattered NLoS paths. The NLoS path angles are randomly perturbed around the nominal BS-UE direction according to the prescribed azimuth and elevation angular spreads.

Second, the SI channel is treated differently because it is governed by near-field mutual coupling, surface-wave and spatial leakage, and other EM interactions between the co-located Tx and Rx arrays. Due to the short Tx-Rx separation and the practical RF-isolation structures used in FD arrays, the SI channel cannot typically be accurately captured by a geometric multipath model. Therefore, instead of relying on an analytical SI model, this work incorporates an experimentally measured SI channel to represent the realistic coupling environment targeted by the proposed SI-aware AS and tri-HBF design.

\subsection{SINR Expressions under HBF}
Based on \eqref{eq:rDi_hbf}, the DL SINR for user $i$ is given by
\begin{equation}
\mathrm{SINR}_{D,i} \hspace{-0.3em}
=\hspace{-0.3em} \frac{p_{D,i}|\bar{\mathbf h}_{D,i}^{H}\mathbf F_D\mathbf b_{D,i}|^{2}}{
\sum_{\tilde i\neq i} p_{D,\tilde i}|\bar{\mathbf h}_{D,i}^{H}\mathbf F_D\mathbf b_{D,\tilde i}|^{2} \hspace{-0.2em}
+ \hspace{-0.2em} \sum_{j=1}^{K_U} p_{U,j}|g_{i,j}|^{2}
\hspace{-0.2em} + \hspace{-0.2em} \sigma_D^{2}}.
\label{eq:sinr_dl_hbf}
\end{equation}

Let $\bar{\mathbf h}_{U,j}$ denote the $j$th column of $\bar{\mathbf H}_U$.
Then, the UL SINR for stream $j$ is expressed as
\begin{equation}
\mathrm{SINR}_{U,j}
\hspace{-0.2em}=\hspace{-0.2em}
\frac{p_{U,j}\,|\mathbf b_{U,j}^{H}\mathbf F_U^{H}\bar{\mathbf h}_{U,j}|^{2}}{
\sum_{k\neq j} p_{U,k}\,|\mathbf b_{U,j}^{H}\mathbf F_U^{H}\bar{\mathbf h}_{U,k}|^{2}
\hspace{-0.1em}+\hspace{-0.1em} I_{U,j}^{\mathrm{SI}}
\hspace{-0.1em}+\hspace{-0.1em} \sigma_{U,j,\mathrm{eff}}^2}.
\label{eq:sinr_ul_hbf}
\end{equation}

Here, the effective post-combining noise variance is 
\begin{equation}
\sigma_{U,j,\mathrm{eff}}^2
\triangleq
\sigma_U^2\,\mathbf b_{U,j}^{H}\mathbf F_U^{H}\mathbf F_U\,\mathbf b_{U,j}
=
\sigma_U^2\|\mathbf F_U\mathbf b_{U,j}\|^2,
\label{eq:noise_ul_hbf}
\end{equation}
and the residual SI power is defined as
\begin{equation}
I_{U,j}^{\mathrm{SI}}
\triangleq
\sum_{i=1}^{K_D} p_{D,i}\,
\big|\mathbf b_{U,j}^{H}\mathbf F_U^{H}\bar{\mathbf H}_{\mathrm{SI}}\mathbf F_D\mathbf b_{D,i}\big|^{2}.
\label{eq:IUj_SI}
\end{equation}

\section{AO-Based Tri-HBF Optimization}\label{sec:methodology}
\subsection{Problem Formulation}
To exploit the design flexibility of the AS-enabled reconfigurable sub-arrays, the proposed tri-HBF design is formulated as a weighted sum-rate (WSR) maximization problem
\begin{subequations}
\label{prob:P0_joint_AS_HBF_detailed}
\begin{align}
\hspace{-0.5em}\max_{\substack{\mathcal X_4=\{\mathbf S_D,\mathbf S_U\},\\
\mathcal X_3=\{\mathbf F_D,\mathbf F_U\},\\
\mathcal X_2=\{\mathbf P_D,\mathbf P_U\},\\
\mathcal X_1=\{\mathbf B_D,\mathbf B_U\}
}} \hspace{-0.5em}
&\begin{aligned}[t]
f(\mathcal X)
\triangleq
&\sum_{i=1}^{K_D}\omega_{D,i}
\log_2\!\left(1+\mathrm{SINR}_{D,i}(\mathcal X)\right)\\
&+
\sum_{j=1}^{K_U}\omega_{U,j}
\log_2\!\left(1+\mathrm{SINR}_{U,j}(\mathcal X)\right)
\end{aligned}
\label{prob:P0_obj}\\
\mathrm{s.t.}\quad
&\mathrm{tr}\!\left(
\mathbf F_D\mathbf B_D\mathbf P_D\mathbf B_D^H\mathbf F_D^H
\right)
\leq P_D^{\mathrm{tot}},
\quad \mathbf P_D\succeq \mathbf 0,
\label{prob:P0_DLpow}\\
&0\leq p_{U,j}\leq P_{U,\max},\quad j=1,\ldots,K_U,
\label{prob:P0_ULpow}\\
&\mathbf S_D\in\{0,1\}^{M_{\mathrm{Tx}}\times M_D},
\quad
\mathbf S_D^T\mathbf S_D=\mathbf I_{M_D},
\label{prob:P0_SD_constr}\\
&\mathbf S_U\in\{0,1\}^{M_{\mathrm{Rx}}\times M_U},
\quad
\mathbf S_U^T\mathbf S_U=\mathbf I_{M_U},
\label{prob:P0_SU_constr}\\
& |\mathcal Q_D\cap \mathcal G_{D,n}|=L_D,\quad n=1,\ldots,N_D, \label{prob:P0_group_D}\\
& |\mathcal Q_U\cap \mathcal G_{U,n}|=L_U,\quad n=1,\ldots,N_U, \label{prob:P0_group_U} \\
&\mathbf F_D\in\mathcal F_D,\quad
\mathbf F_U\in\mathcal F_U .
\label{prob:P0_RFset}
\end{align}
\end{subequations}
Here, the complete optimization-variable set is defined as
\begin{equation}
\mathcal X
\triangleq
\{
\mathbf B_D,\mathbf B_U,
\mathbf P_D,\mathbf P_U,
\mathbf F_D,\mathbf F_U,
\mathbf S_D,\mathbf S_U
\}.
\label{eq:opt_variable_set}
\end{equation}
The objective in \eqref{prob:P0_obj} jointly accounts for the DL and UL spectral efficiencies under AS, HBF, power allocation, and multiple sources of interference. 
The coefficients \(\omega_{D,i}\geq0\) and \(\omega_{U,j}\geq0\) denote the priority weights of DL user \(i\) and UL user \(j\), respectively and determine the relative emphasis on the DL and UL rates in the WSR objective.
In this study, equal importance is assigned to all DL users or all UL users. To further emphasize the UL performance, which is more sensitive to residual SI in FD operation, the UL weights are set \(1.2\) times larger than the DL weights.
\eqref{prob:P0_DLpow} imposes the total DL transmit-power budget at the BS, while \eqref{prob:P0_ULpow} enforces the per-user UL power limits. Constraints \eqref{prob:P0_SD_constr} and \eqref{prob:P0_SU_constr} ensure valid AS matrices.
The group-wise constraints in \eqref{prob:P0_group_D} and \eqref{prob:P0_group_U} preserve the pre-connected sub-array architecture by selecting exactly \(L_D\) and \(L_U\) antenna elements from each predefined Tx and Rx sub-array, respectively. 
Constraint \eqref{prob:P0_RFset} imposes the feasible RF-beamforming sets under the adopted sub-connected constant-modulus HBF architecture.

Problem \eqref{prob:P0_joint_AS_HBF_detailed} is a mixed discrete-continuous non-convex optimization problem. 
The non-convexity arises from the binary AS variables, the constant-modulus RF constraints, and the strong coupling among AS, RF beamforming, BB beamforming, and power allocation in the SINR expressions. 
The use of a realistically measured SI channel further strengthens this coupling because near-field mutual coupling, surface-wave leakage, spatial leakage, and array-dependent SI propagation jointly affect the residual SI terms. 
Therefore, globally solving \eqref{prob:P0_joint_AS_HBF_detailed} is generally intractable, which motivates the AO-based solution developed in this work.
To facilitate AO, the optimization variables are partitioned into four blocks: \(\mathcal X_1=\{\mathbf B_D,\mathbf B_U\}\), \(\mathcal X_2=\{\mathbf P_D,\mathbf P_U\}\), \(\mathcal X_3=\{\mathbf F_D,\mathbf F_U\}\), and \(\mathcal X_4=\{\mathbf S_D,\mathbf S_U\}\). At the \(t\)th AO iteration, these blocks are updated sequentially as \( \mathcal X_1^{(t+1)} \rightarrow \mathcal X_2^{(t+1)} \rightarrow \mathcal X_3^{(t+1)} \rightarrow \mathcal X_4^{(t+1)},\) while the remaining blocks are fixed at their latest values.

\subsection{SI-Aware RZF/MMSE-Based Baseband $\mathcal X_1$ Update}
The MMSE combiner is adopted for the UL BB update because it accounts for the covariance of UL MUI, residual SI, and receiver noise, thereby providing an interference-aware receive filter.
For the DL BB update, an RZF-type precoder is employed to provide a low-complexity and numerically stable balance between desired-signal enhancement and MUI suppression. The additional SI-penalty term further tailors the precoder to FD operation by discouraging beamforming that generates strong SI at the BS receiver.

Given $\mathcal X_2, \mathcal X_3, \mathcal X_4 $, the RF-domain effective channels are
\begin{equation}
\begin{aligned}
\mathbf H_{D,\mathrm{eff}}
&\triangleq 
\mathbf F_D^H\mathbf S_D^T\mathbf H_D^{(\mathrm{full})}
\in\mathbb C^{N_D\times K_D},\\
\mathbf H_{U,\mathrm{eff}}
&\triangleq 
\mathbf F_U^H\mathbf S_U^T\mathbf H_U^{(\mathrm{full})}
\in\mathbb C^{N_U\times K_U},\\
\mathbf H_{\mathrm{SI}}
&\triangleq 
\mathbf F_U^H\mathbf S_U^T
\mathbf H_{\mathrm{SI}}^{(\mathrm{full})}
\mathbf S_D\mathbf F_D
\in\mathbb C^{N_U\times N_D}.
\end{aligned}
\label{eq:effective_channels_RF_domain}
\end{equation}

Let \(\mathbf h_{U,\mathrm{eff},j}\) denote the \(j\)th column of \(\mathbf H_{U,\mathrm{eff}}\). The UL interference-plus-noise covariance for detecting stream \(j\) is
\begin{equation}
\begin{aligned}
\mathbf R_{U,j}\hspace{-0.2em}
\triangleq \hspace{-0.2em}
\sum_{k\neq j}^{K_U} \hspace{-0.1em}
p_{U,k}\mathbf h_{U,\mathrm{eff},k}
\mathbf h_{U,\mathrm{eff},k}^H
\hspace{-0.2em}+\hspace{-0.2em}
\mathbf H_{\mathrm{SI}}\mathbf B_D\mathbf P_D
\mathbf B_D^H\mathbf H_{\mathrm{SI}}^H
\hspace{-0.2em}+ \hspace{-0.2em} \tilde{\sigma}_U\mathbf I_{N_U},
\end{aligned}
\label{eq:Ru_j}
\end{equation}
where \(\mathbf P_D=\mathrm{diag}(\mathbf p_D)\). The term \(\tilde{\sigma}_U\) denotes the RF-domain effective noise variance, including a small diagonal loading for numerical stability. The UL BB combiner is updated as
\begin{equation}
\widetilde{\mathbf b}_{U,j}
=\mathbf R_{U,j}^{-1}\mathbf h_{U,\mathrm{eff},j},
\quad j=1,\ldots,K_U,
\label{eq:BU_mmse_raw}
\end{equation}
followed by antenna-domain normalization
\begin{equation}
\mathbf b_{U,j}
\leftarrow
\frac{\widetilde{\mathbf b}_{U,j}}
{\max\!\left(\|\mathbf F_U\widetilde{\mathbf b}_{U,j}\|_2,\varepsilon\right)},
\quad j=1,\ldots,K_U.
\label{eq:BU_normalize}
\end{equation}
which improves numerical robustness and keeps the effective antenna-domain combiner scale controlled.

For the DL BB precoder, let \(\mathbf h_{D,\mathrm{eff},i}\) denote the \(i\)th column of \(\mathbf H_{D,\mathrm{eff}}\). The weighted DL channel covariance is defined as
\begin{equation}
\mathbf H_{\mathrm w}
\triangleq
\sum_{i=1}^{K_D}
\omega_{D,i}
\mathbf h_{D,\mathrm{eff},i}
\mathbf h_{D,\mathrm{eff},i}^H
\in\mathbb C^{N_D\times N_D}.
\label{eq:H_weighted}
\end{equation}
To account for SI caused by DL transmission, an SI penalty matrix is constructed in the DL RF domain as
\begin{equation}
\mathbf C_{\mathrm{SI}}
\triangleq
\mathbf H_{\mathrm{SI}}^H
\left(
\sum_{j=1}^{K_U}
\omega_{U,j}\mathbf b_{U,j}\mathbf b_{U,j}^H
\right)
\mathbf H_{\mathrm{SI}}
\in\mathbb C^{N_D\times N_D}.
\label{eq:Csi_def}
\end{equation}
The DL BB precoder is updated via an SI-aware RZF rule
\begin{equation}
\mathbf B_D
=
\left(
\mathbf H_{\mathrm w}
+
\alpha\mathbf I_{N_D}
+
\lambda_{\mathrm{SI}}^{\mathrm{BB}}\mathbf C_{\mathrm{SI}}
\right)^{-1}
\mathbf H_{D,\mathrm{eff}},
\label{eq:BD_update_prog}
\end{equation}
where \(\alpha>0\) is the regularization factor and \(\lambda_{\mathrm{SI}}^{\mathrm{BB}}\geq0\) controls the strength of the SI-aware penalty. The DL precoder columns are further normalized in the antenna domain as
\begin{equation}
\mathbf v_i
\triangleq
\mathbf F_D\mathbf b_{D,i},
\;
\mathbf b_{D,i}
\hspace{-0.2em}\leftarrow \hspace{-0.2em}
\frac{\mathbf b_{D,i}}
{\max\!\left(\|\mathbf v_i\|_2,\varepsilon\right)},
\; i=1,\ldots,K_D.
\label{eq:BD_normalize}
\end{equation}
The UL-combiner and DL-precoder updates are repeated for a small fixed number of inner iterations, resulting in an SI-aware and numerically robust BB refresh within each AO iteration.

\subsection{SCA-Based Surrogate for Power Allocation $\mathcal X_2$ Update}
Let \(\mathbf p_D\in\mathbb R_+^{K_D}\) and \(\mathbf p_U\in\mathbb R_+^{K_U}\) denote the DL and UL power-allocation vectors, respectively. Given $\mathcal X_1, \mathcal X_3, \mathcal X_4 $, define the antenna-domain effective precoder and combiner as
\begin{subequations}
\begin{align}
\mathbf V &\triangleq \mathbf F_D\mathbf B_D
=[\mathbf v_1,\ldots,\mathbf v_{K_D}]
\in\mathbb C^{M_D\times K_D},\label{eq:V_W_def} \\
\mathbf W &\triangleq \mathbf F_U\mathbf B_U
=[\mathbf w_1,\ldots,\mathbf w_{K_U}]
\in\mathbb C^{M_U\times K_U}.
\end{align}
\end{subequations}

With fixed beamformers, the SINR denominator and numerator-plus-denominator terms are affine functions of \((\mathbf p_D,\mathbf p_U)\). For DL user \(i\), define
\begin{equation}
\label{eq:GammaD_def}
\Gamma_{D,i}(\mathbf p_D,\mathbf p_U)
\hspace{-0.2em}=\hspace{-0.2em} \sum_{k\neq i}
|\bar{\mathbf h}_{D,i}^H\mathbf v_k|^2p_{D,k}\hspace{-0.2em}+ \hspace{-0.2em} \sum_{j=1}^{K_U}
|g_{i,j}|^2p_{U,j} \hspace{-0.2em} + \hspace{-0.2em} \sigma_D^2,
\end{equation}
\begin{equation}
\label{eq:SigmaD_def}
\Sigma_{D,i}(\mathbf p_D,\mathbf p_U)
=
\Gamma_{D,i}(\mathbf p_D,\mathbf p_U)
+
|\bar{\mathbf h}_{D,i}^H\mathbf v_i|^2p_{D,i}.
\end{equation}

Similarly, for UL user $j$, the interference-plus-noise power and total received power are
\begin{equation}
\begin{aligned}
    \label{eq:GammaU_def}
\Gamma_{U,j}(\mathbf p_D,\mathbf p_U)&=\sum_{k\neq j} |\mathbf w_j^H\bar{\mathbf h}_{U,k}|^2 p_{U,k}
\\&
+\hspace{-0.2em}\sum_{i=1}^{K_D} |\mathbf w_j^H(\bar{\mathbf H}_{\mathrm{SI}}\mathbf v_i)|^2 p_{D,i}\hspace{-0.2em}
+\sigma_U^2\|\mathbf w_j\|_2^2,
\end{aligned}
\end{equation}
\begin{equation}
\label{eq:SigmaU_def}
\Sigma_{U,j}(\mathbf p_D,\mathbf p_U)
\triangleq
\Gamma_{U,j}(\mathbf p_D,\mathbf p_U)+|\mathbf w_j^H\bar{\mathbf h}_{U,j}|^2 p_{U,j}.
\end{equation}
Thus, the rate can be written in difference-of-concave form as
\begin{equation}
\begin{aligned}
R_{D,i}=\log_2(1+\mathrm{SINR}_{D,i})
&\hspace{-0.2em}=\hspace{-0.2em}
\log_2\!\left(\Sigma_{D,i}\right)
\hspace{-0.2em}-\hspace{-0.2em}
\log_2\!\left(\Gamma_{D,i}\right),\\
R_{U,j}=\log_2(1+\mathrm{SINR}_{U,j})
&\hspace{-0.2em}=\hspace{-0.2em}
\log_2\!\left(\Sigma_{U,j}\right)
\hspace{-0.2em}-\hspace{-0.2em}
\log_2\!\left(\Gamma_{U,j}\right).
\end{aligned}
\label{eq:DC_form_rates}
\end{equation}

At AO iteration \(t\), the convex terms \(-\log_2(\Gamma_{D,i})\) and \(-\log_2(\Gamma_{U,j})\) are lower-bounded by their first-order approximations at the current point \((\mathbf p_D^{(t)},\mathbf p_U^{(t)})\). Equivalently, the surrogate rates are written as
\begin{subequations}
\begin{align}
\widetilde R_{D,i}^{(t)}
=&
\log_2\!\left(\Sigma_{D,i}\right)
-\log_2\!\left(\Gamma_{D,i}^{(t)}\right)
-
\frac{\Gamma_{D,i}-\Gamma_{D,i}^{(t)}}
{\ln(2)\Gamma_{D,i}^{(t)}},
\\
\widetilde R_{U,j}^{(t)}
=&
\log_2\!\left(\Sigma_{U,j}\right)
-
\log_2\!\left(\Gamma_{U,j}^{(t)}\right)
-
\frac{\Gamma_{U,j}-\Gamma_{U,j}^{(t)}}
{\ln(2)\Gamma_{U,j}^{(t)}}.
\end{align}
\label{eq:surrogate_rates}
\end{subequations}
where \(\Gamma_{D,i}^{(t)}=\Gamma_{D,i}(\mathbf p_D^{(t)},\mathbf p_U^{(t)})\) and \(\Gamma_{U,j}^{(t)}=\Gamma_{U,j}(\mathbf p_D^{(t)},\mathbf p_U^{(t)})\). The resulting SCA power-allocation subproblem is
\begin{equation}
\label{eq:power_surrogate_problem_prog}
\begin{aligned}
\max_{\mathbf p_D,\mathbf p_U}\quad
&
\sum_{i=1}^{K_D}
\omega_{D,i}\widetilde R_{D,i}^{(t)}
+
\mu_U
\sum_{j=1}^{K_U}
\omega_{U,j}\widetilde R_{U,j}^{(t)}
\\
\mathrm{s.t.}\quad
&
\sum_{i=1}^{K_D}
\|\mathbf F_D\mathbf b_{D,i}\|_2^2p_{D,i}
\leq P_D^{\mathrm{tot}},
\quad
\mathbf p_D\succeq \mathbf 0,\\
&
0\leq p_{U,j}\leq P_{U,\max},
\quad j=1,\ldots,K_U,
\end{aligned}
\end{equation}
where \(\mu_U\) controls the relative emphasis on the UL sum-rate.
The surrogate objective in \eqref{eq:power_surrogate_problem_prog} is concave in \((\mathbf p_D,\mathbf p_U)\), lower-bounds the original WSR objective, and is tight at the current point \((\mathbf p_D^{(t)},\mathbf p_U^{(t)})\). Therefore, solving or improving \eqref{eq:power_surrogate_problem_prog} yields a valid SCA step and produces a nondecreasing WSR sequence when a monotonic acceptance rule is applied.

Since \(\Sigma_{D,i}\), \(\Gamma_{D,i}\), \(\Sigma_{U,j}\), and \(\Gamma_{U,j}\) are affine in \((\mathbf p_D,\mathbf p_U)\), the functions \(\log_2(\Sigma_{D,i})\) and \(\log_2(\Sigma_{U,j})\) are concave, whereas \(-\log_2(\Gamma_{D,i})\) and \(-\log_2(\Gamma_{U,j})\) are convex. The first-order approximation of a convex function gives a global lower bound that is tight at the expansion point. Summing the resulting bounds over all DL and UL users gives the surrogate properties.
The surrogate problem is solved by projected gradient ascent with backtracking
\begin{subequations}
\begin{align}
\mathbf p_D
&\leftarrow
\Pi_{\mathcal P_D}
\left(
\mathbf p_D+\eta\nabla_{\mathbf p_D}\widetilde f^{(t)}
\right),\\
\mathbf p_U
&\leftarrow
\Pi_{\mathcal P_U}
\left(
\mathbf p_U+\eta\nabla_{\mathbf p_U}\widetilde f^{(t)}
\right),
\end{align}
\label{eq:PGA_power}
\end{subequations}
where \(\eta\) is selected by backtracking and
\begin{equation}
\begin{aligned}
\mathcal P_D
&\triangleq
\left\{
\mathbf p_D\succeq \mathbf 0:
\sum_i c_{D,i}p_{D,i}
\leq P_D^{\mathrm{tot}}
\right\},\\
\mathcal P_U
&\triangleq
\left\{
\mathbf p_U:
0\leq p_{U,j}\leq P_{U,\max},
\; j=1,\ldots,K_U
\right\},
\end{aligned}
\label{eq:power_feasible_sets}
\end{equation}
with \(c_{D,i}=\|\mathbf F_D\mathbf b_{D,i}\|_2^2\). A monotonic ascent check is applied after each projected-gradient step.
The projection onto the weighted DL power-feasible set is given by
\begin{equation}
\label{eq:proj_weighted_budget}
\Pi_{\mathcal P_D}(\mathbf y)=\arg\min_{\mathbf x\succeq \mathbf 0}
\|\mathbf x-\mathbf y\|_2^2
\quad
\mathrm{s.t.}
\,
\sum_{i} c_{D,i}x_i
\leq P_D^{\mathrm{tot}},
\end{equation}
where \(\mathbf y\) denotes the tentative DL power vector after the gradient-ascent step, and \(\mathbf x\) is the projected feasible power vector. The optimal projection has a water-filling-like form
\begin{equation}
x_i=\max\{y_i-\lambda c_{D,i},0\},
\end{equation}
where \(\lambda\geq0\) is the Lagrange multiplier associated with the weighted DL power constraint. If the constraint is inactive, then \(\lambda=0\); otherwise, \(\lambda\) is chosen such that $\sum_i c_{D,i}x_i=P_D^{\mathrm{tot}}$.
In practice, \(\lambda\) can be efficiently obtained by a one-dimensional bisection search. The UL projection is obtained by simple element-wise clipping due to the independent per-user UL power constraints, i.e.,
\([\Pi_{\mathcal P_U}(\mathbf y_U)]_j=\min\{\max\{y_{U,j},0\},P_{U,\max}\}, \quad j=1,\ldots,K_U \).

\subsection{Phase-Domain PGA for RF Beamformer \(\mathcal X_3\) Update}
Given \(\mathcal X_1\), \(\mathcal X_2\), and \(\mathcal X_4\), the RF beamformers are updated over the feasible sets \(\mathcal F_D\) and \(\mathcal F_U\) using phase-domain PGA. Under the adopted sub-connected constant-modulus architecture, the DL RF precoder and UL RF combiner are parameterized by phase vectors \(\boldsymbol\theta_D\in\mathbb R^{M_D}\) and \(\boldsymbol\theta_U\in\mathbb R^{M_U}\), respectively, such that \(\mathbf F_D(\boldsymbol\theta_D)\in\mathcal F_D\) and \(\mathbf F_U(\boldsymbol\theta_U)\in\mathcal F_U\).
After the AS stage, each DL RF chain is connected to \(L_D\) selected Tx antenna elements, and each UL RF chain is connected to \(L_U\) selected Rx antenna elements within its corresponding sub-array branch.
 With group-wise ordering of the selected antennas, the nonzero entries of the RF beamformers are
\begin{equation}
\label{eq:FD_theta_param_prog}
[\mathbf F_D(\boldsymbol\theta_D)]_{m,n}
=
\begin{cases}
\dfrac{1}{\sqrt{L_D}}\exp(j\theta_{D,m}), 
& m\in\mathcal I_{D,n},\\
0, & \text{otherwise},
\end{cases}
\end{equation}
\begin{equation}
\label{eq:FU_theta_param_prog}
[\mathbf F_U(\boldsymbol\theta_U)]_{m,n}
=
\begin{cases}
\dfrac{1}{\sqrt{L_U}}\exp(j\theta_{U,m}), 
& m\in\mathcal I_{U,n},\\
0, & \text{otherwise},
\end{cases}
\end{equation}
where \(\mathcal I_{D,n}\triangleq \mathcal Q_D\cap \mathcal G_{D,n},
\) and \(\mathcal I_{U,n}\triangleq \mathcal Q_U\cap \mathcal G_{U,n},\)
denote the index sets of the selected Tx and Rx antennas connected to the \(n\)th DL and UL RF chains, respectively.
Let \(f(\mathbf F_D,\mathbf F_U)\) denote the WSR objective in \eqref{prob:P0_joint_AS_HBF_detailed} with all other blocks fixed. 
The DL and UL phase vectors are updated by phase-domain
PGA as
\begin{equation}\label{eq:PGA_theta_DL_UL_prog}
\boldsymbol\theta_D^{(t+1)}=\boldsymbol\theta_D^{(t)}
+\eta_D^{(t)}
\nabla_{\boldsymbol\theta_D} f,\;\;
\boldsymbol\theta_U^{(t+1)}
=\boldsymbol\theta_U^{(t)}
+\eta_U^{(t)}
\nabla_{\boldsymbol\theta_U} f,
\end{equation}
followed by RF-beamformer reconstruction
\begin{equation}
\label{eq:RF_reconstruction_prog}
\mathbf F_D^{(t+1)}
=
\mathbf F_D(\boldsymbol\theta_D^{(t+1)}),
\qquad
\mathbf F_U^{(t+1)}
=
\mathbf F_U(\boldsymbol\theta_U^{(t+1)}),
\end{equation}
where \(\eta_D^{(t)}\) and \(\eta_U^{(t)}\) are the DL and UL step sizes, respectively. 

The phase gradient is approximated using symmetric finite differences
\begin{equation}
\label{eq:finite_diff_grad}
[\nabla_{\boldsymbol\theta} f]_k
\approx
\frac{
f(\boldsymbol\theta+\epsilon\mathbf e_k)
-
f(\boldsymbol\theta-\epsilon\mathbf e_k)
}{2\epsilon}, \; \boldsymbol\theta\in\{\boldsymbol\theta_D,\boldsymbol\theta_U\}
\end{equation}
where \(\epsilon>0\) is a small perturbation step and \(\mathbf e_k\) is the \(k\)th canonical basis vector. The resulting gradient is normalized, and a backtracking line search is applied to ensure monotonic ascent for each accepted RF update.

Unlike the digital BB update, the RF-beamformer subproblem does not admit a tractable closed-form solution due to the coupled WSR objective, the sub-connected structure, and the constant-modulus constraints. Therefore, the RF block is optimized numerically in the phase domain. The adopted phase parameterization guarantees feasibility after each update, while the backtracking rule ensures that accepted RF updates do not decrease the WSR objective.

\subsection{SI-Aware Utility with WSR Acceptance for AS $\mathcal X_4$ Update}
Given \((\mathcal X_1,\mathcal X_2,\mathcal X_3)\), the AS update is a discrete combinatorial problem. Exhaustive joint search over all feasible Tx/Rx subsets is computationally intractable, especially under the group-wise sub-connected constraint. Therefore, a low-complexity SI-aware utility is adopted to update \((\mathbf S_D,\mathbf S_U)\). The key idea is to evaluate separate per-antenna utilities on the Tx and Rx sides, followed by top-\(L_D\) and top-\(L_U\) selection within each predefined Tx/Rx sub-array group. 

Given the current beamformers, the corresponding full-array embedded DL precoder and UL combiner vectors are defined as
\begin{subequations}
\begin{align}
\tilde{\mathbf v}_i
&\triangleq
\mathbf S_D\mathbf F_D\mathbf b_{D,i}
\in\mathbb C^{M_{\mathrm{Tx}}\times 1},
\\
\tilde{\mathbf w}_j
&\triangleq
\mathbf S_U\mathbf F_U\mathbf b_{U,j}
\in\mathbb C^{M_{\mathrm{Rx}}\times 1}.
\end{align}
\end{subequations}

For candidate Tx antenna \(m\) and candidate Rx antenna \(n\), the desired-signal utilities are defined as
\begin{subequations}
\begin{align}
\label{eq:tx_usig_prog}
u_{\mathrm{sig}}^{D}(m)
&=
\sum_{i=1}^{K_D}
\omega_{D,i}p_{D,i}
\left|
[\mathbf h_{D,i}^{(\mathrm{full})}]_m
\right|^2,\\
\label{eq:rx_usig_prog}
u_{\mathrm{sig}}^{U}(n)
&=
\sum_{j=1}^{K_U}
\omega_{U,j}p_{U,j}
\left|
[\mathbf H_U^{(\mathrm{full})}]_{n,j}
\right|^2.
\end{align}
\end{subequations}
which quantify the weighted desired-signal contributions of candidate Tx and Rx antennas under the current DL and UL power allocations, respectively.

In FD operation, AS needs to balance desired-signal enhancement and SI suppression under the current HBF state. Therefore, the proposed utility incorporates both a desired-channel gain term and an SI-leakage penalty. The Tx- and Rx-side SI leakage proxies are defined as
\begin{subequations}
\begin{align}
\label{eq:tx_usi_state}
u_{\mathrm{si}}^{D}(m)&=\sum_{j=1}^{K_U}
\mu_U\omega_{U,j}
\left|
\tilde{\mathbf w}_j^H
[\mathbf H_{\mathrm{SI}}^{(\mathrm{full})}]_{:,m}
\right|^2,\\
\label{eq:rx_usi_state}
u_{\mathrm{si}}^{U}(n)
&=
\sum_{i=1}^{K_D}
p_{D,i}
\left|
[\mathbf H_{\mathrm{SI}}^{(\mathrm{full})}\tilde{\mathbf v}_i]_n
\right|^2.
\end{align}
\end{subequations}
Here, \(u_{\mathrm{si}}^{D}(m)\) quantifies the SI leakage induced by activating Tx antenna \(m\) toward the current UL combining directions, while \(u_{\mathrm{si}}^{U}(n)\) quantifies the SI power received at Rx antenna \(n\) under the current DL precoding state.
Hence, the selection rule favours antennas that provide strong desired-signal support while avoiding those that cause excessive SI leakage.

To balance the relative magnitudes of the desired signal and SI terms, each quantity is normalized by its empirical mean
\begin{subequations}
\begin{align}
\mathcal N_{\mathrm{Tx}}(x_m)
&\triangleq
\frac{x_m}
{\max\left\{\frac{1}{M_{\mathrm{Tx}}}\sum_{m'=1}^{M_{\mathrm{Tx}}}x_{m'},\,\varepsilon\right\}},\\
\mathcal N_{\mathrm{Rx}}(y_n)
&\triangleq
\frac{y_n}
{\max\left\{\frac{1}{M_{\mathrm{Rx}}}\sum_{n'=1}^{M_{\mathrm{Rx}}}y_{n'},\,\varepsilon\right\}},
\end{align}
\label{eq:normalization_ops}
\end{subequations}
where \(\varepsilon>0\) is a small regularization constant. The final Tx- and Rx- side utilities are then given by
\begin{subequations}\label{eq:util_final_prog}
\begin{align}
u_D(m)
&=
\mathcal N_{\mathrm{Tx}}\!\left(u_{\mathrm{sig}}^{D}(m)\right)
-
\lambda_{\mathrm{SI}}
\mathcal N_{\mathrm{Tx}}\!\left(u_{\mathrm{si}}^{D}(m)\right),
\label{eq:util_tx_final_prog}\\
u_U(n)
&=
\mathcal N_{\mathrm{Rx}}\!\left(u_{\mathrm{sig}}^{U}(n)\right)
-
\lambda_{\mathrm{SI}}
\mathcal N_{\mathrm{Rx}}\!\left(u_{\mathrm{si}}^{U}(n)\right),
\label{eq:util_rx_final_prog}
\end{align}
\end{subequations}
for \(m=1,\ldots,M_{\mathrm{Tx}}\) and \(n=1,\ldots,M_{\mathrm{Rx}}\), where \(\lambda_{\mathrm{SI}}\geq0\) controls the desired-signal/SI tradeoff. 
The candidate \(\mathbf S_D\) and \(\mathbf S_U\) are then formed by selecting the top-\(L_D\) Tx antennas and top-\(L_U\) Rx antennas within each predefined Tx/Rx sub-array group, thereby preserving the pre-connected sub-array constraint.
To improve robustness with respect to the WSR objective, the AS update is further combined with a monotonic acceptance rule. For a candidate selection \((\mathbf S_D',\mathbf S_U')\), the algorithm performs local re-optimization, including RF phase-matching initialization, SI-aware BB refresh, optional SCA-based power refinement, and optional short RF-PGA refinement. The update is accepted only if \(f(\mathcal X') > f(\mathcal X)+\delta\), where \(\delta>0\) is a small acceptance tolerance. This prevents detrimental discrete selection changes and ensures that accepted AS updates improve the locally re-optimized WSR.
Overall, the proposed AS update balances desired-channel enhancement and SI suppression under the current HBF state. The desired-signal terms preserve BS-UE channel gain, while the SI-penalty terms exploit the SI channel \(\mathbf H_{\mathrm{SI}}^{(\mathrm{full})}\) and the current beamformer  \(\{\mathbf v_i\}\) and \(\{\mathbf w_j\}\). Thus, \eqref{eq:tx_usi_state} and \eqref{eq:rx_usi_state} capture both antenna-level coupling and beamformed SI, making the selection rule more consistent with the FD WSR objective.

Additionally, the proposed separable AS utility avoids the combinatorial complexity of joint Tx/Rx subset search. A pairwise Tx/Rx scoring strategy requires at least \(\mathcal O(M_{\mathrm{Tx}}M_{\mathrm{Rx}})\) operations, while exhaustive group-wise joint selection scales as
\( \mathcal O\!\left(
\binom{M_{D,s}}{L_D}^{N_D}
\binom{M_{U,s}}{L_U}^{N_U}
\right)\), which becomes prohibitive for large-scale arrays. In contrast, the proposed update only requires per-antenna utility evaluation followed by group-wise top-\(L_D\) and top-\(L_U\) selection, resulting in a polynomial-complexity AS update compatible with the adopted sub-connected RF architecture.
Compared with PSO-based joint optimization, the proposed AO framework is more structured and scalable. A PSO-based method requires \(N_{\rm p}\) WSR evaluations per swarm iteration over \(T_{\rm PSO}\) iterations, with complexity \(\mathcal O\!\left(T_{\rm PSO}N_{\rm p}C_{\rm eval}\right)\), where \(C_{\rm eval}\) is the cost of constructing feasible AS/HBF/power variables and evaluating the WSR. When jointly optimizing AS, RF phases, BB variables, and power allocation, PSO operates over a high-dimensional mixed binary-continuous search space, with a significant number of particles and iterations.
Compared with conventional HBF without AS, the proposed AO scheme introduces additional polynomial complexity from AS and SI-aware power optimization; however, the group-wise AS structure and strategic utility design keep this overhead manageable while exploiting selection diversity to avoid strongly coupled Tx-Rx pairs.

Since the beamformer and power allocation updates depend on the selected effective channels, a group-wise SI-only rule is first applied to obtain an initialization \((\mathbf S_D^{(0)},\mathbf S_U^{(0)})\) by selecting antenna subsets with the lowest Tx-Rx leakage within each predefined sub-array. The RF beamformers \((\mathbf F_D^{(0)},\mathbf F_U^{(0)})\) are then initialized through phase matching within the selected sub-arrays, followed by an SI-aware initialization of the BB precoder/combiner \((\mathbf B_D^{(0)},\mathbf B_U^{(0)})\). A DL soft-start power strategy is further adopted by initializing \(\mathbf p_D\) to a fraction of \(P_D^{\mathrm{tot}}\), which mitigates excessive early-stage SI to the UL receiver. 

\begin{algorithm}[!t]
\caption{AO-Based Algorithm for Joint Tri-HBF}
\label{alg:ao_joint_ashbfpl}
\begin{algorithmic}[1]
\State \textbf{Input:} DL/UL user locations, user weights $\{\omega_{D,i}\}$ and $\{\omega_{U,j}\}$, power limits $P_D^{\mathrm{tot}}$ and $P_{U,\max}$
\State \textbf{Initialization:} Obtain $(\mathbf S_D^{(0)},\mathbf S_U^{(0)})$ by the group-wise SI-only rule.  Initialize $(\mathbf F_D^{(0)},\mathbf F_U^{(0)})$, $(\mathbf B_D^{(0)},\mathbf B_U^{(0)})$, $(\mathbf p_D^{(0)},\mathbf p_U^{(0)})$. Set $t\gets 0$ and compute $f^{(0)}$ from \eqref{prob:P0_obj}

\For{$t=0,1,\ldots,T_{\max}-1$}

    \State \parbox[t]{0.9\linewidth}{\textbf{Block 1:} Update $\mathcal X_1=\{\mathbf B_D,\mathbf B_U\}$ by the SI-aware alternating MMSE/RZF refresh in \eqref{eq:BU_normalize} and \eqref{eq:BD_normalize}.}

    \State \parbox[t]{0.9\linewidth}{\textbf{Block 2:} Construct the SCA surrogate in \eqref{eq:power_surrogate_problem_prog};}
    \State \parbox[t]{0.9\linewidth}{Update $\mathcal X_2=\{\mathbf P_D,\mathbf P_U\}$ by \eqref{eq:PGA_power} over \eqref{eq:power_feasible_sets}, with the DL projection given by \eqref{eq:proj_weighted_budget}.}

    \State \parbox[t]{0.9\linewidth}{\textbf{Block 3:} Compute the phase gradients by \eqref{eq:finite_diff_grad};}
    \State \parbox[t]{0.9\linewidth}{Update $(\boldsymbol\theta_D,\boldsymbol\theta_U)$ by PGA using \eqref{eq:PGA_theta_DL_UL_prog};}
    \State Reconstruct $\mathcal X_3=\{\mathbf F_D,\mathbf F_U\}$ via \eqref{eq:FD_theta_param_prog} and \eqref{eq:FU_theta_param_prog}.

    \State \parbox[t]{0.9\linewidth}{\textbf{Block 4:} Generate AS candidate \(\mathcal X_4'=\{\mathbf S_D',\mathbf S_U'\}\) using the utility \eqref{eq:util_final_prog} and group-wise selection;}
    \State \parbox[t]{0.9\linewidth}{Accept the candidate only if the WSR increases after local re-optimization and parameter fine-tuning.}

    \If{$|f^{(t+1)}-f^{(t)}|/\max\{1,|f^{(t)}|\}\leq\epsilon_{\mathrm{AO}}$}
        \State \textbf{break}
    \EndIf
\EndFor
\State \Return $\hat{\mathcal X}$
\end{algorithmic}
\end{algorithm}

\section{Illustrative Results} \label{Illustrative}
\subsection{Measured Large-Scale Array SI Channel}
\begin{figure}[!t]
\centerline{\includegraphics[width=0.8\linewidth]{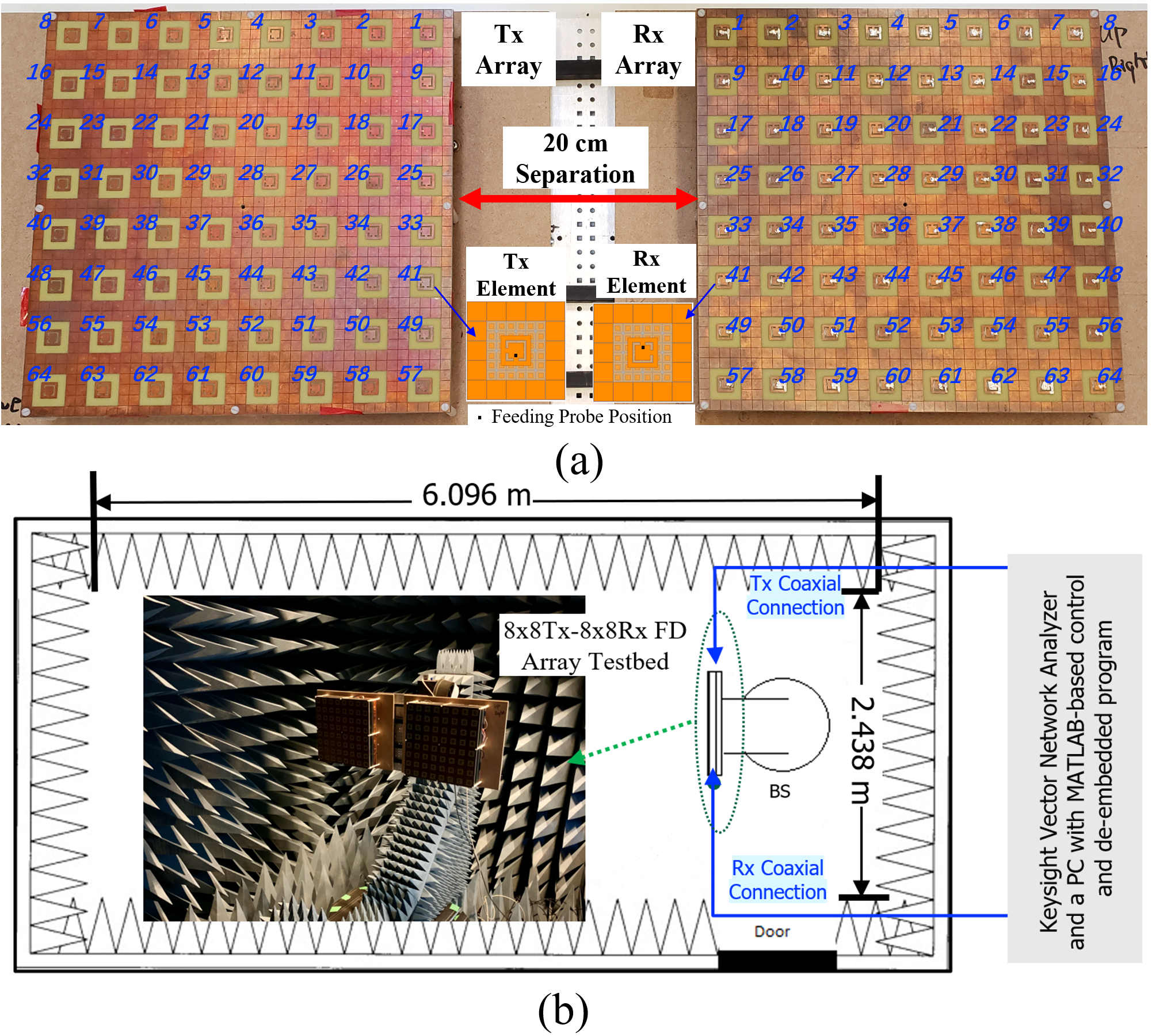}}
\vspace{-0.2cm}
\caption{(a) 8$\times$8Tx-8$\times$8Rx FD mMIMO Testbed with element indices, (b) SI-channel measurement setup in the anechoic chamber.}
\label{SI_channel_meas}
\vspace{-0.2cm}
\end{figure}

As shown in Fig.~\ref{SI_channel_meas}, an \(8\times 8\)Tx-\(8\times8\)Rx antenna-array prototype with our previously developed cross-polarized Tx/Rx antenna elements operating in the frequency range of 3.35-3.6 GHz \cite{YG_ACCESS_1,gong2026subarray} is employed to obtain measurement-based SI channels for FD experiments. 
Within each array, a center-to-center inter-element spacing of 4.0 cm (\(\approx 0.5\) wavelength), is adopted to avoid spatial aliasing and grating-lobe formation.
All elements in each array are fabricated on a common FR-4 dielectric substrate with a shared ground plane. Each Tx/Rx array has a physical aperture of \(32~\mathrm{cm}\times 32~\mathrm{cm}\) with 64 Tx/Rx antenna elements, and the Tx and Rx arrays are separated by 20 cm.
To characterize the mutual coupling between the Tx and Rx antenna elements and obtain realistic SI channel data, over-the-air (OTA) FD measurements were conducted in the anechoic chamber, which suppresses external reflections and provides a controlled measurement environment. The measured S-parameters were imported into MATLAB for post-processing, where the insertion losses introduced by the measurement setup were de-embedded to extract the intrinsic antenna-level Tx-Rx coupling.
The calibrated SI channel data at sample 3.5 GHz were then assembled into a complex-valued SI matrix, denoted by \(\mathbf H_{\mathrm{SI}}^{(\mathrm{full})}\). To ensure consistency between the measurement and simulation models, a strict geometric mapping and Tx-Rx element pairing were maintained between the physical prototype and the simulated array.

\subsection{Simulation Setup}
The BS concurrently serves \(K_D=4\) DL and \( K_U=4\) UL users. 
For each Monte-Carlo realization, the DL and UL users are randomly dropped within a service region located \(30\) to \(50\) m from the array, with an azimuthal span of \(\pm 60^\circ\) around boresight. The BS is assumed to be \(10\) m higher than the user height. A total of \(100\) Monte Carlo drops are used for performance evaluation.
The BS-user channels are generated using a narrowband Rician multipath model. Specifically, each user channel contains a LoS component with probability \(P_{\rm LoS}=0.5\) and angularly spread NLoS components, with path-loss exponent \(\eta=3.5\), Rician factor \(K_{\rm R}=10\) dB, and \(6\) propagation paths. 
An additional \(45\) dB Tx-Rx isolation is assumed from the combined RF- and/or BB-domain SIC stages. 
Since this study focuses on BS-side optimization, the UL-user leakage to the DL users is assumed to be sufficiently separated and isolated.
The operating bandwidth is \(20\) MHz centred at \(3.5\) GHz, and the Rx noise power is calculated with a thermal noise density of \(-174\) dBm/Hz and a noise figure of \(5\) dB.  
A sub-connected HBF architecture is adopted.
Unless otherwise specified, \(N_D=N_U=4\) RF chains are used. The Tx and Rx arrays are partitioned into predefined sequential sub-array groups, where the \(n\)th RF chain is connected to the \(n\)th group. Accordingly, the \(n\)th DL RF chain is associated with Tx antenna indices \((n-1)M_{D,s}+1,\ldots,nM_{D,s}\), while the \(n\)th UL RF chain is associated with Rx antenna indices \((n-1)M_{U,s}+1,\ldots,nM_{U,s}\). 

 For performance comparison, several representative AS/HBF baselines are considered. 
1) \textbf{Random Sel.}: randomly selects feasible antennas within each predefined sub-array group. 
2) \textbf{Fixed Sel.}: activates the first \(L_D\) Tx and \(L_U\) Rx elements in each DL/UL sub-array. 
3) \textbf{Desired-only Sel.}: selects antennas according to the BS--UE desired-channel gain. 
4) \textbf{SI-only Sel.}: selects Tx/Rx antennas to minimize SI coupling channel. 
5) \textbf{Greedy-SI Sel.}: selects antennas using beamforming-aware SI-leakage proxies under the current HBF state.
For fairness, all baselines satisfy the same group-wise sub-connected AS constraints and are evaluated under identical RF-chain, power, channel, and user settings. Among them, Random Sel., Desired-only Sel., SI-only Sel., and Greedy-SI Sel. perform adaptive antenna selection based on different FD mMIMO metrics, whereas Fixed Sel. reduces to conventional HBF with a fixed sub-array configuration.

\subsection{Number of Selected Antenna versus the Channel Quality}
\begin{figure}[!t]
\centerline{\includegraphics[width=0.95\linewidth]{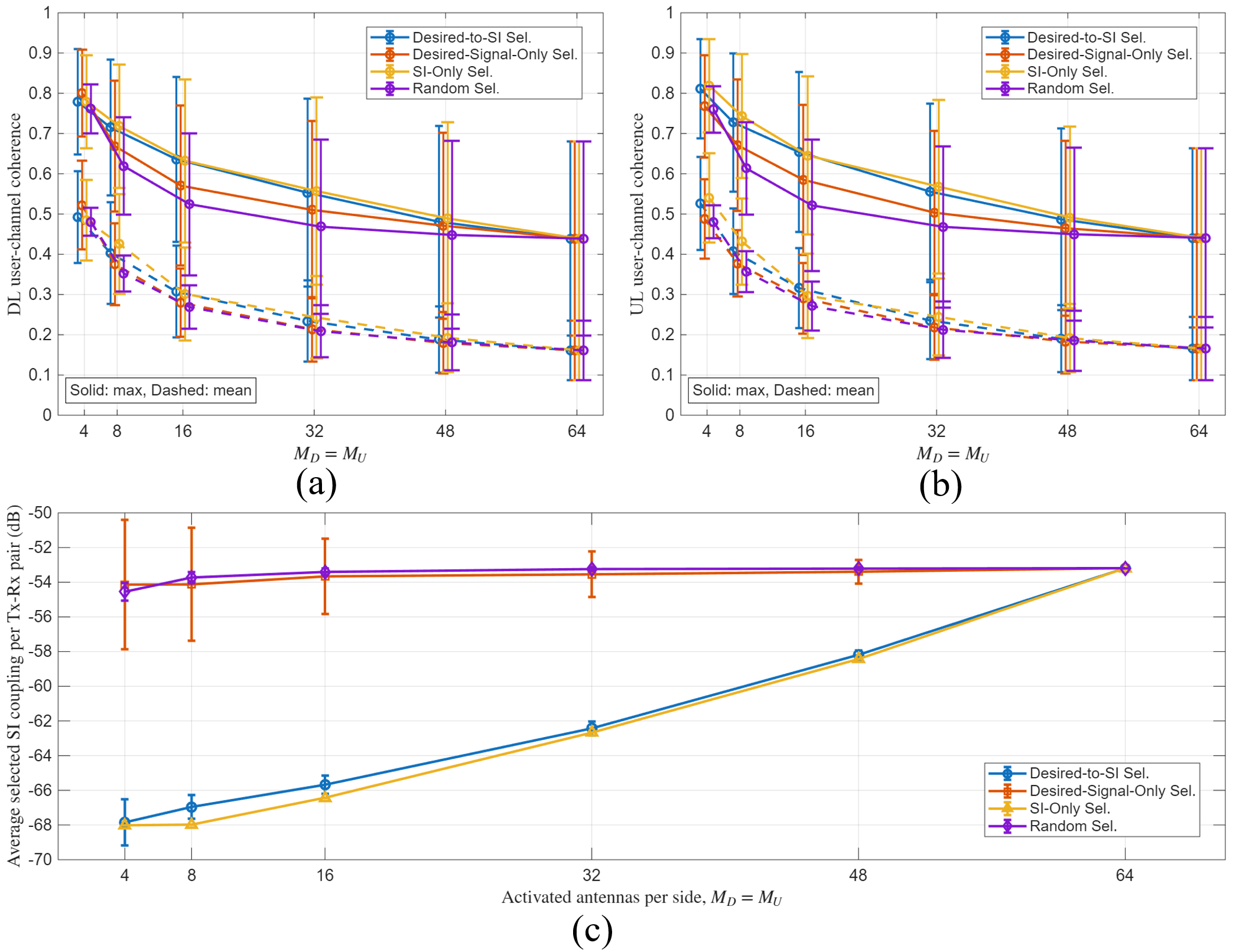}}
\vspace{-0.2cm}
\caption{Desired channel coherence versus the selected number of antenna elements for (a) DL and (b) UL channel. (c) Normalized average selected SI coupling versus the number of selected antennas.}
\vspace{-0.2cm}
\label{channel_sel}
\end{figure}

To study the potential of AS in FD mMIMO, Fig.~\ref{channel_sel} evaluates the selected BS-user channel separability and the SI tradeoff versus the number of activated antenna elements. 
For user-channel separability, the pairwise coherence is computed from the normalized user-channel vectors as $|\frac{\mathbf h_i^H\mathbf h_k} {\|\mathbf h_i\|_2\|\mathbf h_k\|_2}|,$ and $ i\neq k$.
Fig.~\ref{channel_sel}~(a) and~(b) present the maximum and mean pairwise coherence of the selected DL and UL user-channel matrices, respectively. 
The maximum coherence captures the most highly correlated user pair and therefore reflects the worst-case user separability, whereas the mean coherence measures the average similarity among all user-channel pairs.
Both metrics generally decrease as \(M_D=M_U\) increases, indicating that activating more antenna elements provides higher RF spatial resolution and improves user-channel distinguishability. 
This enhanced separability increases the potential of the subsequent beamforming stages to suppress MUI. However, the coherence reduction becomes progressively smaller as more elements are activated, suggesting diminishing marginal improvement in user separability for large selected subsets.

To quantify the FD performance of the selected antenna subsets, Fig.~\ref{channel_sel} (c) presents the normalized average selected SI coupling and its distribution versus the number of activated elements. 
When fewer antennas are activated, SI-aware selection schemes, including the SI-only and desired-to-SI schemes, provide a notable improvement in Tx-Rx isolation by favoring antennas with lower SI leakage or a higher desired-channel gain relative to SI leakage.
For instance, reducing \(M_D=M_U\) from 64 to 4 yields \(15.0\) dB reduction in the average SI coupling. 
This improvement is mainly due to the increased selection diversity, which allows the algorithm to avoid strongly coupled Tx-Rx antenna pairs and effectively exploit the spatial non-uniformity of the SI channel.
In contrast, the random and desired-signal-only selection schemes exhibit average SI coupling levels close to that of the full-array case, since they do not explicitly explore the SI channel during selection.

Overall, increasing the number of activated antennas improves user separability, whereas selecting fewer but more favorable antennas provides a stronger desired-signal/SI tradeoff. Therefore, in FD mMIMO, full-array activation is not necessarily optimal; instead, an intermediate activated subset can better balance MUI suppression and SI mitigation.

\subsection{Convergence Behaviour of the Proposed AO scheme}
\begin{figure}[!t]
\centerline{\includegraphics[width=0.8\linewidth]{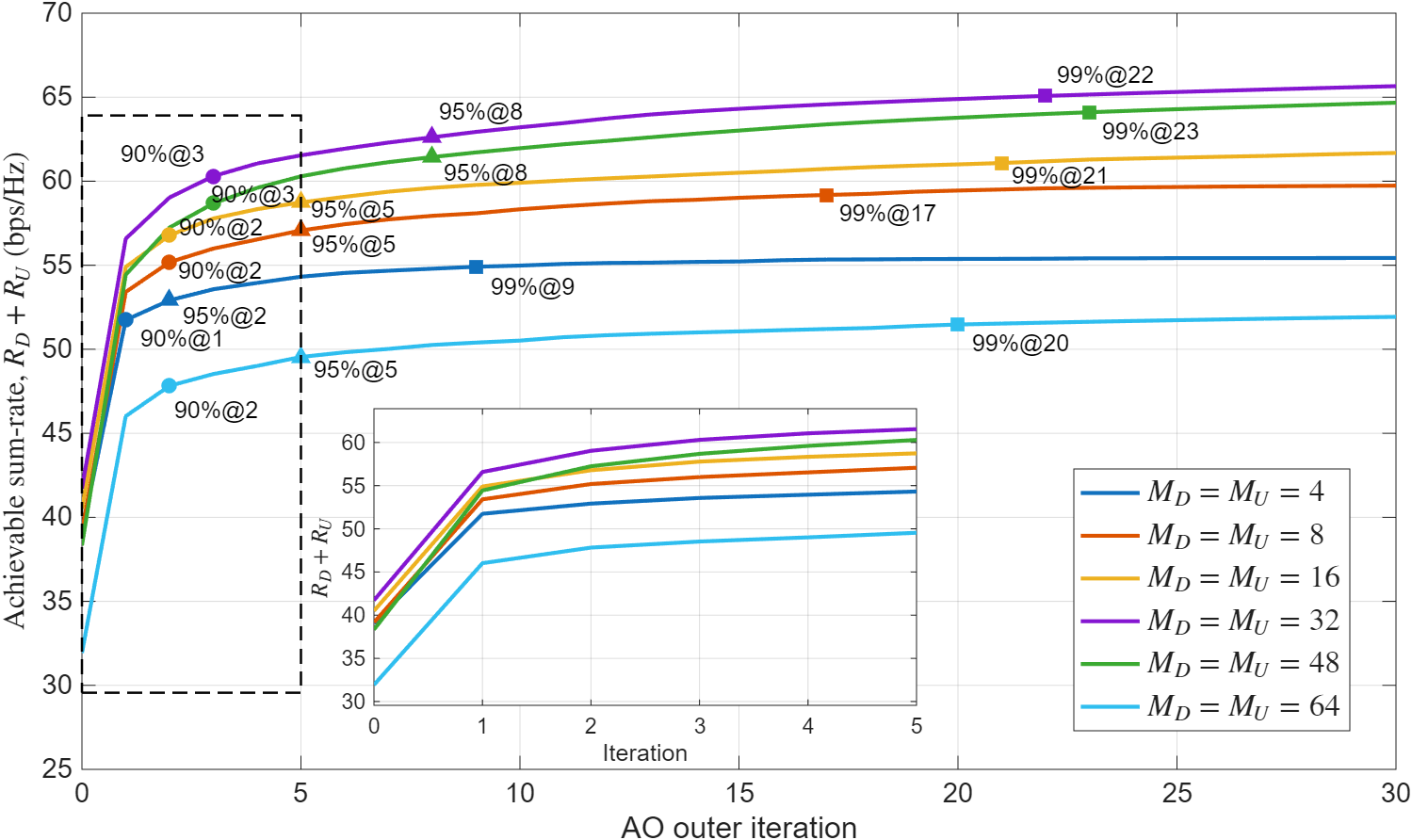}}
\vspace{-0.2cm}
\caption{Proposed AO scheme convergence behaviour versus the number of activated antennas and $N_D=N_U=4$.}
\vspace{-0.2cm}
\label{Convergence}
\end{figure}

\subsubsection{Convergence Versus the Number of Activated Antennas}
Fig.~\ref{Convergence} presents the convergence behaviour of the proposed AO-based tri-HBF optimization scheme in terms of the Monte-Carlo-averaged achievable sum-rate under different numbers of activated Tx/Rx antenna elements. 
The RF-chain and user settings are fixed as \(N_D=N_U=K_D=K_U=4\), while \(M_D=M_U\) is varied to evaluate the impact of the activated antenna dimension on both convergence speed and achievable-rate performance.
The proposed algorithm exhibits rapid and robust convergence for all tested antenna configurations. Specifically, the AO procedure achieves \(90\%\), \(95\%\), and \(99\%\) of the final achievable sum-rate within the first 3, 8, and 22 outer iterations, respectively. 
A slightly slower convergence trend is observed as the number of activated elements increases, since larger selected arrays introduce more RF phase variables and a higher-dimensional effective channel/SI structure. Nevertheless, doubling the number of activated antennas typically requires only 3 additional outer iterations to reach a comparable convergence level, indicating that the proposed AO framework remains computationally stable as the selected array size increases.

The results also illustrate that activating more antennas does not necessarily lead to a higher FD mMIMO achievable rate. With the fixed RF-chain setting \(N_D=N_U=4\), the configuration \(M_D=M_U=32\) achieves the highest average sum-rate. Further increasing the number of activated antennas to \(M_D=M_U=48\) leads to rate saturation or degradation, revealing the tradeoff between improved spatial resolution and increased SI coupling under limited RF-chain resources.

\begin{figure}[!t]
\centerline{\includegraphics[width=0.8\linewidth]{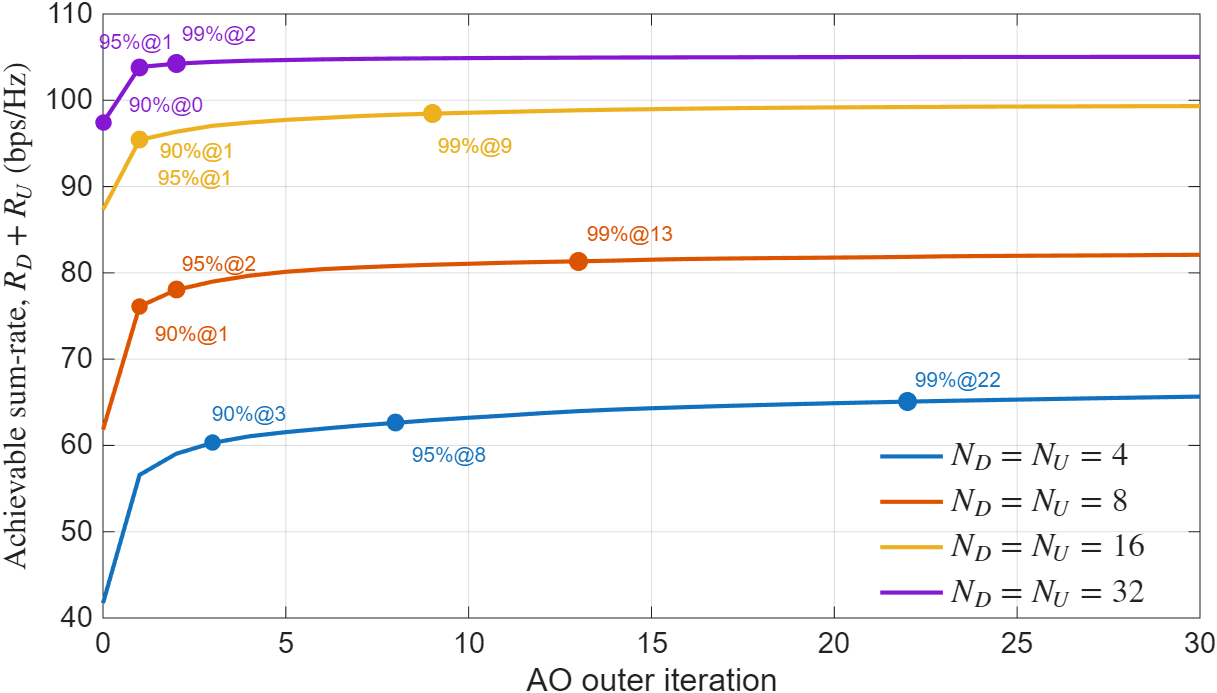}}
\vspace{-0.2cm}
\caption{Proposed AO scheme convergence behaviour versus the available number of RF chains and $M_D=M_U=32$.}
\vspace{-0.2cm}
\label{Convergence_RF}
\end{figure}

\subsubsection{Convergence Versus the Number of RF Chains} Fig.~\ref{Convergence_RF} presents the convergence behavior of the proposed AO-based tri-HBF scheme under different numbers of DL/UL RF chains, with fixed activated antenna and user settings \(M_D=M_U=32\) and \(K_D=K_U=4\). \(N_D=N_U\) is varied to evaluate the impact of the RF-chain dimension on both convergence speed and achievable-rate performance.
The proposed AO-based tri-HBF scheme maintains robust convergence across all tested RF-chain setups. As the number of RF chains increases, fewer outer iterations are required to reach a comparable convergence level, indicating that the enlarged RF-chain dimension alleviates the RF-domain bottleneck and improves the effectiveness of the joint RF/BB optimization.
The final achievable sum-rate increases as \(N_D=N_U\) increases, since more RF chains provide a higher-dimensional effective BB channel and more spatial DoFs for DL precoding, UL combining, MUI suppression, and SI-aware beamforming. 
However, the marginal gain decreases as the number of RF chains becomes large. 
For example, doubling \(N_D=N_U\) from 4 to 8 yields an approximately \(25\%\) sum-rate improvement, whereas doubling it from 16 to 32 provides only about a \(5\%\) improvement. 
This saturation occurs because, for a fixed number of activated antennas, increasing the number of RF chains reduces the number of antenna elements connected to each sub-array. Consequently, the additional RF-chain DoFs provide progressively smaller beamforming gains, and the performance becomes increasingly limited by the channel condition, residual SI, and finite user separability rather than by the RF-chain dimension alone.

\subsection{Power Efficiency and Sum-Rate Versus Active Antennas}
\begin{figure}[!t]
\centerline{\includegraphics[width=0.8\linewidth]{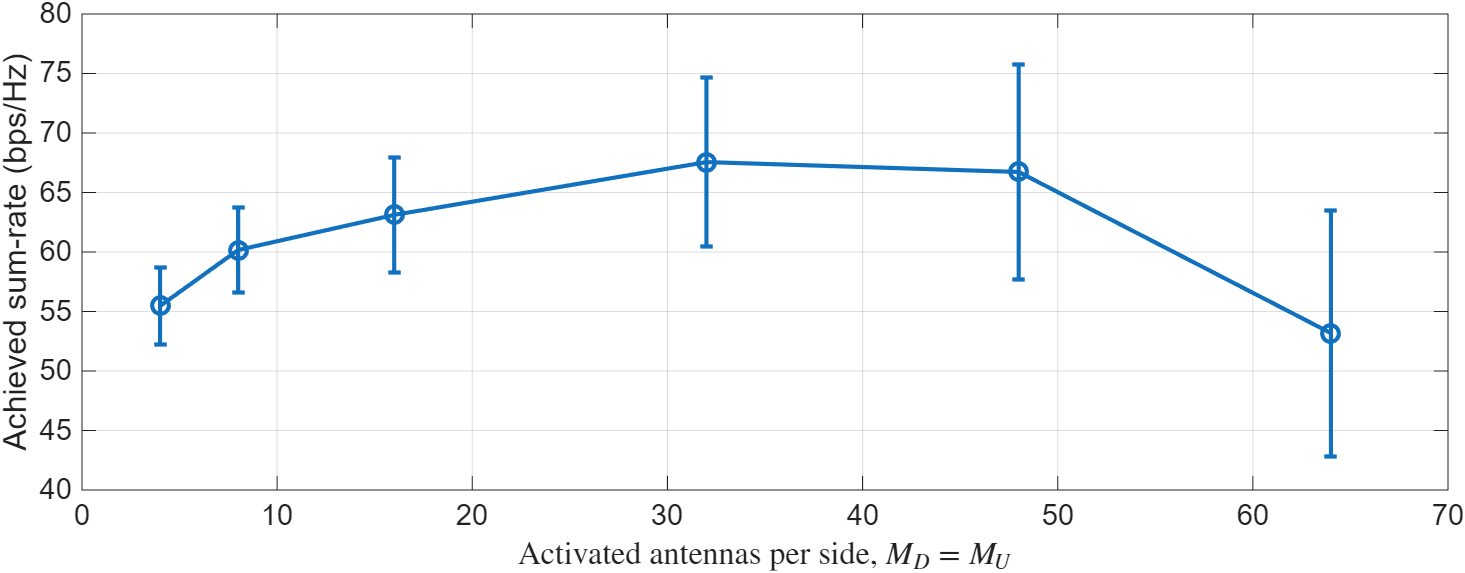}}
\vspace{-0.1cm}
\caption{Monte Carlo evaluation of the AO-based tri-HBF: average achieved and distribution sum-rate versus the number of activated antenna elements.}
\label{Rate_distribution}
\vspace{-0.2cm}
\end{figure}

Fig.~\ref{Rate_distribution} presents the Monte-Carlo-averaged sum-rate and its distribution under different numbers of activated DL and UL antenna elements, with fixed RF-chain and user settings \(N_D=N_U=K_D=K_U=4\). 
As \(M_D=M_U\) increases from 4 to 32, the average sum-rate improves from \(55.5\) bps/Hz to \(67.6\) bps/Hz, corresponding to a \(21.8\%\) gain. Beyond \(M_D=M_U=32\), the rate improvement saturates, indicating diminishing returns from further antenna activation.
This trend is consistent with the channel-level observations on user-channel coherence and the desired-to-SI tradeoff.
Compared with the best-performing configuration \(M_D=M_U=32\), full activation results in an approximately \(21.3\%\) reduction in the sum-rate. 
This degradation highlights the adverse impact of excessive SI coupling and reduced selection diversity under limited RF-chain resources.
The rate distribution further confirms this tradeoff. With a smaller selected subset, the AS mechanism can more effectively exclude unfavorable Tx/Rx elements, resulting in a more robust rate distribution across user-location realizations. 
For example, when \(M_D=M_U=4\), the best-to-worst sum-rate variation is about \(3.2\) bps/Hz, corresponding to approximately \(5.8\%\) of the average rate. In contrast, when \(M_D=M_U=48\), the variation increases to about \(9.0\) bps/Hz, or approximately \(13.6\%\) of the average rate. 

Therefore, although increasing \(M_D\) and \(M_U\) improves array gain and user separability, the corresponding gain gradually saturates once the number of activated elements exceeds a certain range. Meanwhile, activating more Tx/Rx elements also introduces additional unfavourable SI coupling paths in FD and reduces the sub-array reconfigurable freedom.
Therefore, under a limited number of RF chains, strategic antenna selection enables the tri-HBF architecture to better balance spatial multiplexing, MUI suppression, and SI mitigation, thereby improving FD mMIMO performance.

Partial antenna activation is also attractive from a practical power-efficiency perspective. Although the above results are obtained under a fixed transmit-power constraint, practical power consumption also includes RF-chain circuits, phase shifters, switches, bias/control circuits, BB processing, and power amplifier DC power. A representative power model with \(M\) active antennas can be written as
\begin{equation}
P_{\rm tot}=
P_{\rm PA}^{\rm DC}+
N_{RF}P_{\rm RFC}
+M(P_{\rm PS}+P_{\rm SW})
+P_{\rm BB}
+P_{\rm fix},
\end{equation}
where \(P_{\rm PA}^{\rm DC}\) is the amplifier DC power, \(P_{\rm RFC}\) is the per-RF-chain circuit power, \(P_{\rm PS}\) and \(P_{\rm SW}\) are the per-active-path phase-shifter and switching/control powers, \(P_{\rm BB}\) is the BB-processing power, and \(P_{\rm fix}\) denotes fixed circuit overhead.

For a fixed RF-chain setup, reducing \(M\) decreases the number of active antenna RF paths and hence reduces the associated phase-shifter, switching, biasing, and feeding-network power.
For example, assuming \(P_{\rm PS}=30\) mW and \(P_{\rm SW}=1\) mW per antenna path \cite{mendez2016hybrid}, reducing the activated antennas from \(M_D=M_U=64\) to \(M_D=M_U=32\) saves
\(2(64-32)(P_{\rm PS}+P_{\rm SW})
=
64\times31~{\rm mW}
\approx 1.98~{\rm W}\)
from the phase-shifter and switching/control paths alone. Additional savings may be obtained from reduced bias/control power, lower feeding-network loss, and relaxed RF dynamic-range requirements.

The power amplifier's DC power depends on the adopted RF front-end architecture. 
Element-level amplifier implementation is generally favorable for active mMIMO arrays because the amplifier is placed close to each antenna element, thereby avoiding high-power signal distribution through lossy splitter/phase-shifter feeding networks. 
This architecture distributes the transmit power over multiple low-power amplifier branches, mitigates insertion-loss-induced backoff, and enables inactive antennas to be power-gated. With an element-level amplifier scheme, deactivating antennas can directly shut down the corresponding PA paths and considerably reduce both PA bias power and associated RF front-end circuit power. 
On the other hand, if the amplifier is implemented at the RF-chain level, reducing \(M_D\) does not directly reduce the PA count, but it can still reduce the required PA output power by lowering the insertion loss of the analog feeding network, where the amplifier DC power can be modelled as
\(P_{\rm PA}^{\rm DC}=
\frac{P_{\rm ant}10^{L_{\rm ins}/10}}{\eta_{\rm PA}}\),
where \(\eta_{\rm PA}\) is the amplifier efficiency, \(P_{\rm ant}\) is the required antenna-port power, and \(L_{\rm ins}\) is the insertion loss of the analog feeding network in dB. For example, an \(8\) dB insertion loss corresponds to a linear power factor of \(10^{8/10}\approx 6.3\), implying that the amplifier must deliver approximately \(6.3\) times more RF power before the lossy feeding network to maintain the same antenna-port power. Therefore, selective antenna activation can improve practical power efficiency by reducing active amplifier/RF paths in element-level amplifier architectures and by mitigating feeding-network loss and unnecessary RF-path power consumption in RF-chain-level PA architectures.

Therefore, partial antenna activation can improve not only the achievable rate but also the practical rate-per-watt efficiency. By selecting the favorable antenna elements, the proposed AS-aided tri-HBF scheme preserves most of the useful spatial gain while reducing unnecessary active RF paths, feeding-network loss, and SI-induced inefficiency.

\subsection{Achieved Rate of the Proposed AS-Based Tri-HBF Scheme}
\begin{figure}[!t]
\centerline{\includegraphics[width=0.9\linewidth]{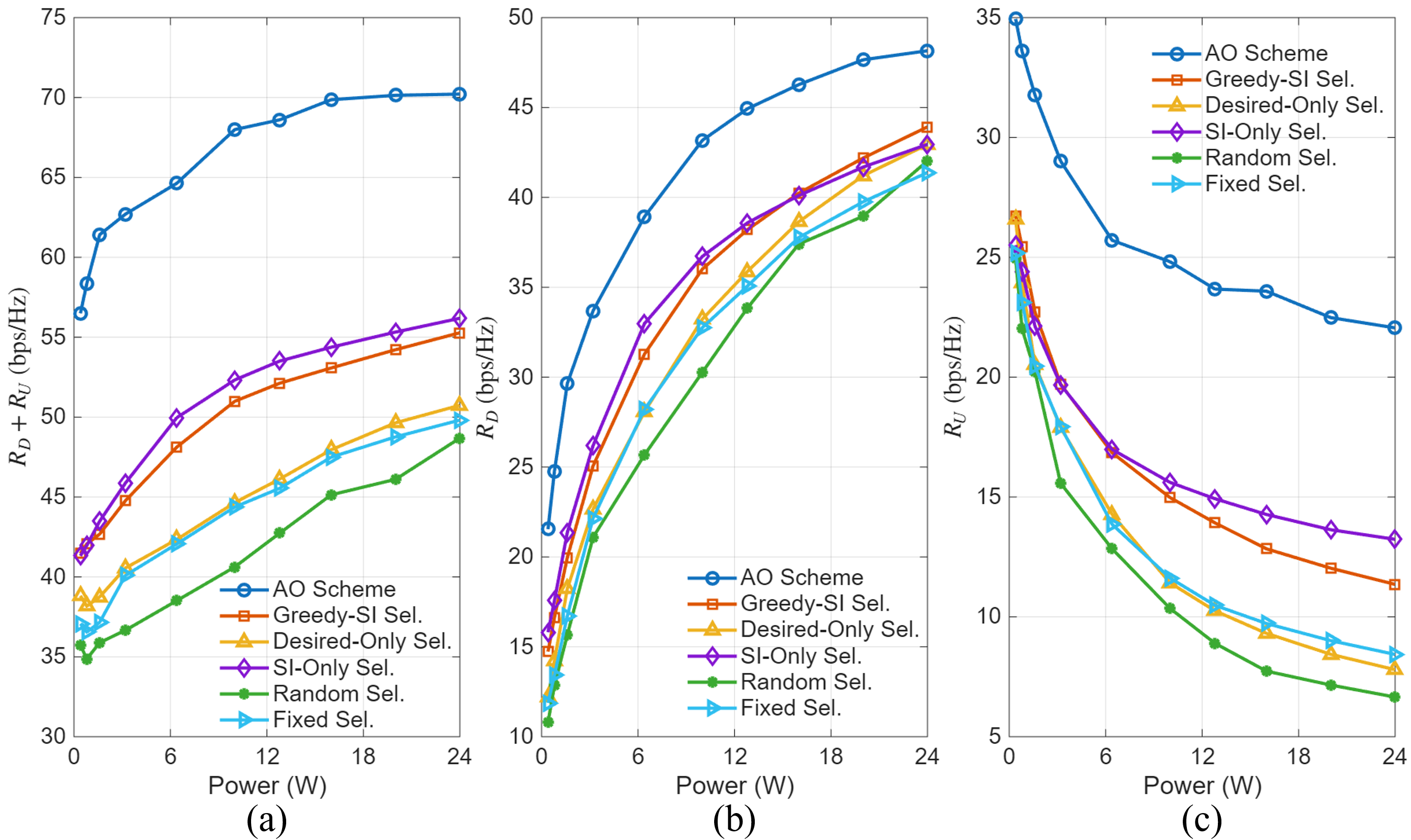}}
\vspace{-0.2cm}
\caption{Average achievable-rate performance of various antenna selection schemes with tri-HBF optimization versus the DL transmit-power limit with \(P_{U,\max}=0.2\)W: (a) sum-rate, (b) DL rate, and (c) UL rate.}
\vspace{-0.2cm}
\label{Rate_DP2_2}
\end{figure}

To illustrate the effectiveness of the proposed tri-HBF optimization over conventional HBF schemes with random, fixed, or non-joint AS, Fig.~\ref{Rate_DP2_2} presents the average achievable rate versus the maximum BS DL transmit power. 
The proposed AS-enabled reconfigurable sub-array provides a notable improvement in the achievable sum-rate. 
Compared with the average performance of the five baseline schemes, the proposed scheme achieves an average gain of $20.0$~bps/Hz, corresponding to a relative rate improvement of $45.1\%$. Across the considered BS transmit-power samples, the rate improvement ranges from $34.7\%$ to $55.1\%$.
Among the baselines, the SI-only and greedy-SI selection schemes achieve the strongest performance, which confirms that SI is a dominant bottleneck in FD mMIMO systems. 
In contrast, random selection gives the lowest rate because it does not effectively exploit the spatial structure of either the desired BS-UE channels or the SI channel. 
Instead, it statistically averages the SI and desired-link effects over the entire Tx/Rx array without targeted antenna selection. 
The fixed selection scheme, which activates a complete $1\times 8$ row of the sub-array, shows performance close to that of the desired-signal-only selection scheme. This is mainly because the selected row preserves a relatively large effective aperture, thereby maintaining RF-domain beam resolution.
The reconfigurable sub-array structure improves both DL and UL rates. Specifically, the proposed scheme achieves average DL and UL rate improvements of $36.9\%$ and $82.9\%$, respectively. These results demonstrate that the proposed joint AS-based tri-HBF scheme can effectively mitigate MUI and SI simultaneously, thereby improving both DL and UL transmission performance in FD mMIMO systems.

\begin{figure}[!t]
\centerline{\includegraphics[width=0.9\linewidth]{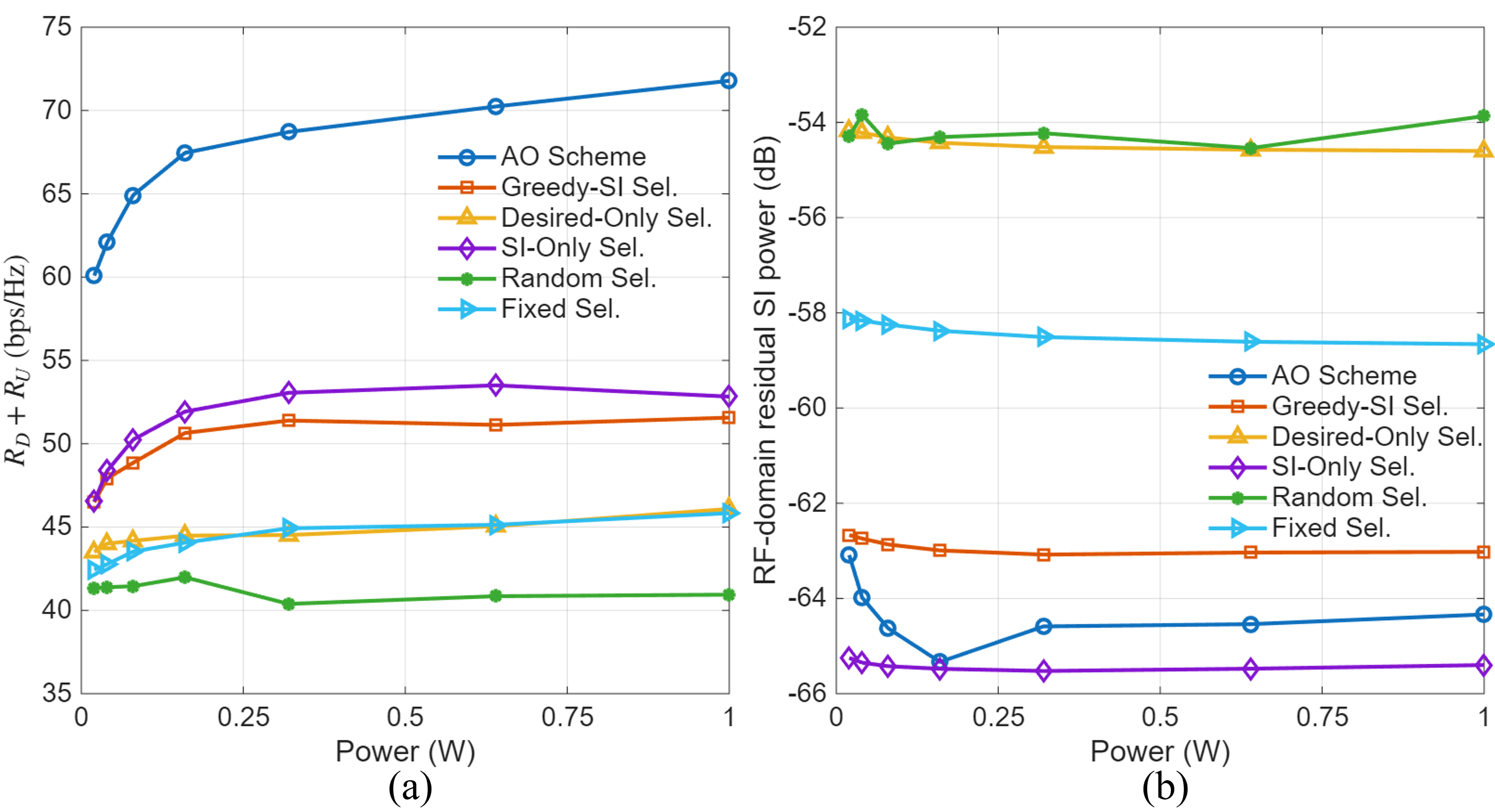}}
\vspace{-0.1cm}
\caption{(a) Average achievable sum rate performance and (b) average RF beam-level SI of various antenna selection schemes with tri-HBF optimization versus the UL per-UE transmit-power limit, with \(P_D^{\mathrm{tot}}=10\)W.}
\vspace{-0.2cm}
\label{Rate_UP}
\end{figure}

A similar trend is observed when sweeping the per-UL-user transmit power, as shown in Fig.~\ref{Rate_UP}. Across the considered UL transmit-power samples, the proposed scheme achieves an average sum-rate gain of $20.4$~bps/Hz, corresponding to a relative improvement of $44.1\%$. 
The achieved sum-rate increases consistently as the power increases, indicating that the proposed AO-based tri-HBF scheme can effectively exploit the increased UL signal power while jointly mitigating SI and MUI.
In contrast, the baseline schemes saturate earlier as the UL transmit power increases, suggesting that they become interference-limited due to residual MUI, SI, and the inherent DL-UL rate tradeoff in FD operation.

Moreover, Fig.~\ref{Rate_UP}~(b) shows a substantial improvement in beam-level isolation, measured by the average RF-domain SI coupling per DL-UL beam pair. The proposed tri-HBF optimization achieves a beam-level SI level close to that of the SI-only selection scheme, with an average gap less than \(1.1\) dB. 
Compared with the desired-signal-only and random selection schemes, the proposed method improves the average beam-level SI isolation by more than \(10\) dB. This confirms that the proposed design not only preserves the desired channel gain but also effectively suppresses the SI coupling after RF beamforming. As a result, the SI power observed at the UL RF chains is significantly reduced, helping alleviate receiver dynamic-range limitations and potential RF-chain distortion.

\section{Conclusions}\label{Conclusions}
This paper proposed an AS-aided tri-HBF scheme with reconfigurable sub-arrays for FD mMIMO systems. An AO framework was developed to jointly optimize the reconfigurable sub-array structure and the HBF variables.
To account for realistic SI behavior, an experimentally measured SI channel obtained from an \(8\times8\) Tx and \(8\times8\) Rx FD antenna-array prototype was incorporated into the evaluation framework. The illustrative results showed that selective antenna activation can outperform full-array activation, achieving a \(21.3\%\) higher average sum-rate and a more consistent performance distribution across user realizations, while also offering potential power-efficiency benefits through reduced active RF-path usage. The empirical analysis further characterized the impact of the number of activated antennas on achievable rate, user-channel coherence, and desired-to-SI gain, revealing the tradeoff between spatial multiplexing capability and SI mitigation.
The proposed AO-based scheme exhibited fast convergence across different activated-antenna and RF-chain configurations. Compared with various selection baselines, it achieved a \(45.1\%\) improvement in average sum-rate, with average DL and UL rate gains of \(36.9\%\) and \(82.9\%\), respectively. Moreover, beam-level SI isolation exceeding \(63\) dB was achieved. These results confirm that the proposed AS-aided tri-HBF design effectively balances desired-signal enhancement, MUI suppression, and SI mitigation in FD mMIMO operation.

\bibliographystyle{IEEEtran}
\bibliography{bib_journal} 

@article{marzetta2010noncooperative,
  title={Noncooperative cellular wireless with unlimited numbers of base station antennas},
  author={Marzetta, Thomas L},
  journal={IEEE Trans. Wireless Commun.},
  volume={9},
  number={11},
  pages={3590--3600},
  year={2010},
  publisher={IEEE}
}

@article{zhang2016full,
  title={{Full-duplex wireless communications: Challenges, solutions, and future research directions}},
  author={Zhang, Zhongshan and Long, Keping and Vasilakos, Athanasios V and Hanzo, Lajos},
  journal={Proc. IEEE},
  volume={104},
  number={7},
  pages={1369--1409},
  year={2016},
  publisher={IEEE}
}

@article{farahani2010mutual,
  title={Mutual coupling reduction in patch antenna arrays using a {UC-EBG} superstrate},
  author={Farahani, Hossein Sarbandi and Veysi, Mehdi and Kamyab, Manouchehr and Tadjalli, Alireza},
  journal={IEEE Antennas Wireless Propag. Lett.},
  volume={9},
  pages={57--59},
  year={2010},
  publisher={IEEE}
}

@article{duarte2013design,
  title={Design and characterization of a full-duplex multiantenna system for {WiFi} networks},
  author={Duarte, Melissa and others},
  journal={IEEE Trans. Veh. Technol.},
  volume={63},
  number={3},
  pages={1160--1177},
  year={2014},
  publisher={IEEE}
}

@article{rusek2012scaling,
  title={{Scaling up MIMO: Opportunities and challenges with very large arrays}},
  author={Rusek, Fredrik and others},
  journal={IEEE Signal Process. Mag.},
  volume={30},
  number={1},
  pages={40--60},
  year={2013},
  publisher={IEEE}
}

@article{sabharwal2014band,
  title={In-band full-duplex wireless: {C}hallenges and opportunities},
  author={Sabharwal, Ashutosh and Schniter, Philip and Guo, Dongning and Bliss, Daniel W and Rangarajan, Sampath and Wichman, Risto},
  journal={IEEE J. Sel. Areas Commun.},
  volume={32},
  number={9},
  pages={1637--1652},
  year={2014},
  publisher={IEEE}
}

@ARTICLE{YG_ACCESS_1,
  author={Gong, Yuanzhe and Mahmood, Mobeen and Morawski, Robert and Le-Ngoc, Tho},
  journal={IEEE Access}, 
  title={Dual-Layer Metamaterial Rectangular Antenna Arrays for In-Band Full-Duplex Massive {MIMO}}, 
  year={2023},
  volume={11},
  number={},
  pages={135708-135727},
  doi={10.1109/ACCESS.2023.3337823}}

@ARTICLE{YG_ACCESS_2,
  author={Gong, Yuanzhe and Morawski, Robert and Le-Ngoc, Tho},
  journal={IEEE Access}, 
  title={Metamaterial Absorber Structure for {Tx-Rx} Antenna Isolation Improvement in Full-Duplex Massive {MIMO}}, 
  year={2024},
  volume={12},
  number={},
  pages={64571-64588},
  doi={10.1109/ACCESS.2024.3396862}}

@article{debaillie2014analog,
  title={{Analog/RF solutions enabling compact full-duplex radios}},
  author={Debaillie, Bj{\"o}rn and others},
  journal={IEEE J. Sel. Areas Commun.},
  volume={32},
  number={9},
  pages={1662--1673},
  year={2014},
  publisher={IEEE}
}

@article{nawaz2017dual,
  title={Dual-polarized, differential fed microstrip patch antennas with very high interport isolation for full-duplex communication},
  author={Nawaz, Haq and Tekin, Ibrahim},
  journal={IEEE Trans. Antennas Propag.},
  volume={65},
  number={12},
  pages={7355--7360},
  year={2017},
  publisher={IEEE}
}

@article{dadgarpour2016mutual,
  title={Mutual coupling reduction in dielectric resonator antennas using metasurface shield for {60-GHz MIMO} systems},
  author={Dadgarpour, Abdolmehdi and Zarghooni, Behnam and Virdee, Bal S and Denidni, Tayeb A and Kishk, Ahmed A},
  journal={IEEE Antennas Wireless Propag. Lett.},
  volume={16},
  pages={477--480},
  year={2017},
  publisher={IEEE}
}

@article{ahmed2015all,
  title={All-digital self-interference cancellation technique for full-duplex systems},
  author={Ahmed, Elsayed and Eltawil, Ahmed M},
  journal={IEEE Trans. Wireless Commun.},
  volume={14},
  number={7},
  pages={3519--3532},
  year={2015},
  publisher={IEEE}
}

@TECHREPORT{Report_5G_Macro_PL_Rel_17,
	AUTHOR =        {{3GPP TR 36.931}},
	TITLE =         {{LTE};
	Evolved Universal Terrestrial Radio Access ({E-UTRA});
	Radio Frequency ({RF}) requirements for {LTE} Pico Node {B}},
	NUMBER =        {Ver. 17.0.0},
	MONTH =         {Apr.},
	YEAR  =         {2022},
}

@article{mahmood2024achieving,
  title={Achieving capacity gains in practical full-duplex massive {MIMO} systems: a multi-objective optimization approach using hybrid beamforming},
  author={Mahmood, Mobeen and Koc, Asil and Morawski, Robert and Le-Ngoc, Tho},
  journal={IEEE Open J. Commun. Soc.},
  volume={5},
  pages={2268--2286},
  year={2024},
  publisher={IEEE}
}

@article{yang2015efficient,
  title={Efficient full-duplex relaying with joint antenna-relay selection and self-interference suppression},
  author={Yang, Kun and Cui, Hongyu and Song, Lingyang and Li, Yonghui},
  journal={IEEE Trans. Wireless Commun.},
  volume={14},
  number={7},
  pages={3991--4005},
  year={2015},
  publisher={IEEE}
}

@article{fidan2018performance,
  title={Performance of transceiver antenna selection in two way full-duplex relay networks over Rayleigh fading channels},
  author={Fidan, Efendi and Kucur, O{\u{g}}uz},
  journal={IEEE Trans. Veh. Technol.},
  volume={67},
  number={7},
  pages={5909--5921},
  year={2018},
  publisher={IEEE}
}

@article{zhu2022antenna,
  title={Antenna selection for full-duplex distributed massive {MIMO} via the elite preservation genetic algorithm},
  author={Zhu, Pengcheng and Sheng, Zheng and Bao, Jialong and Li, Jiamin},
  journal={IEEE Commun. Lett.},
  volume={26},
  number={4},
  pages={922--926},
  year={2022},
  publisher={IEEE}
}

@article{wilson2017antenna,
  title={Antenna selection for full-duplex {MIMO} two-way communication systems},
  author={Wilson-Nunn, Daniel G and Chaaban, Anas and Sezgin, Aydin and Alouini, Mohamed-Slim},
  journal={IEEE Commun. Lett.},
  volume={21},
  number={6},
  pages={1373--1376},
  year={2017},
  publisher={IEEE}
}

@article{jang2016antenna,
  title={Antenna selection schemes in bidirectional full-duplex {MIMO} systems},
  author={Jang, Seokju and Ahn, Minki and Lee, Hoon and Lee, Inkyu},
  journal={IEEE Trans. Veh. Technol.},
  volume={65},
  number={12},
  pages={10097--10100},
  year={2016},
  publisher={IEEE}
}

@article{gao2015massive,
  title={Massive {MIMO} in real propagation environments: {Do} all antennas contribute equally?},
  author={Gao, Xiang and Edfors, Ove and Tufvesson, Fredrik and Larsson, Erik G},
  journal={IEEE Trans. Commun.},
  volume={63},
  number={11},
  pages={3917--3928},
  year={2015},
  publisher={IEEE}
}

@article{ding2024movable,
  title={Movable antenna-aided secure full-duplex multi-user communications},
  author={Ding, Jingze and Zhou, Zijian and Jiao, Bingli},
  journal={IEEE Trans. Wireless Commun.},
    year={2025},
  volume={24},
  number={3},
  pages={2389-2403},
  publisher={IEEE}
}

@article{liu2023joint,
  title={Joint transmit and receive beamforming design in full-duplex integrated sensing and communications},
  author={Liu, Ziang and Aditya, Sundar and Li, Hongyu and Clerckx, Bruno},
  journal={IEEE J. Sel. Areas Commun.},
  volume={41},
  number={9},
  pages={2907--2919},
  year={2023},
  publisher={IEEE}
}

@article{nguyen2020spectral,
  title={On the spectral and energy efficiencies of full-duplex cell-free massive {MIMO}},
  author={Nguyen, Hieu V and others},
  journal={IEEE J. Sel. Areas Commun.},
  volume={38},
  number={8},
  pages={1698--1718},
  year={2020},
  publisher={IEEE}
}

@book{le2026half,
  title={Half-Duplex/Full-Duplex MU-mMIMO: Metamaterial-Based Large-Scale Array \& Hybrid Beamforming Design},
  author={Le-Ngoc, Tho and Gong, Yuanzhe and Mahmood, Mobeen},
  year={2026},
  publisher={Springer Nature}
}

@article{le2024full,
  title={Full-duplex in massive multiple-input multiple-output},
  author={Le-Ngoc, Tho and Gong, Yuanzhe and Mahmood, Mobeen and Koc, Asil and Morawski, Robert and Griffiths, James Gary and Guillemette, Philippe and Zaid, Jamal and Wang, Peiwei},
  journal={IEEE Open J. Veh. Technol.},
  volume={5},
  pages={560--576},
  year={2024},
  publisher={IEEE}
}

@article{hong2022frequency,
  title={Frequency-domain {RF} self-interference cancellation for in-band full-duplex communications},
  author={Hong, Zhihong Hunter and others}, 
  journal={IEEE Trans. Wireless Commun.},
  volume={22},
  number={4},
  pages={2352--2363},
  year={2023},
  publisher={IEEE}
}

@article{li2025joint,
  title={Joint discrete antenna positioning and beamforming optimization in movable antenna enabled full-duplex {ISAC} networks},
  author={Li, Zhendong and others},
  journal={IEEE Trans. Wireless Commun.},
  volume={25},
  pages={7220--7234},
  year={2026},
  publisher={IEEE}
}

@article{hong2025fluid,
  title={Fluid antenna system-assisted self-interference cancellation for in-band full duplex communications},
  author={Hong, Hanjiang and others},
  journal={IEEE Trans. Wireless Commun.},
  year={2026},
  volume={25},
  number={},
  pages={7476-7489},
  publisher={IEEE}
}

@article{skouroumounis2023fluid,
  title={Fluid antenna-aided full duplex communications: A macroscopic point-of-view},
  author={Skouroumounis, Christodoulos and Krikidis, Ioannis},
  journal={IEEE J. Sel. Areas Commun.},
  volume={41},
  number={9},
  pages={2879--2892},
  year={2023},
  publisher={IEEE}
}

@article{shan2026fluid,
  title={Fluid Antenna-Aided Semi--Full--Duplex System With Non-Orthogonal Multiple Access},
  author={Shan, Yibo and others},
  journal={IEEE Wireless Commun. Lett.},
  volume={15},
  pages={2594--2598},
  year={2026},
  publisher={IEEE}
}

@article{tang2025full,
  title={{Full-duplex FAS-assisted base station for ISAC}},
  author={Tang, Boyi and others},
  journal={IEEE Trans. Wireless Commun.},
  year={2026},
  volume={25},
  number={},
  pages={2922-2938},
  publisher={IEEE}
}

@article{abdullah2022low,
  title={Low-complexity antenna selection and discrete phase-shifts design in IRS-assisted multiuser massive {MIMO} networks},
  author={Abdullah, Zaid and Chen, Gaojie and Lambotharan, Sangarapillai and Chambers, Jonathon A},
  journal={IEEE Trans. Veh. Technol.},
  volume={71},
  number={4},
  pages={3980--3994},
  year={2022},
  publisher={IEEE}
}

@article{gao2017massive,
  title={Massive {MIMO} antenna selection: Switching architectures, capacity bounds, and optimal antenna selection algorithms},
  author={Gao, Yuan and Vinck, Han and Kaiser, Thomas},
  journal={IEEE Trans. Signal Process.},
  volume={66},
  number={5},
  pages={1346--1360},
  year={2018},
  publisher={IEEE}
}

@article{wu2023ris,
  title={{RIS}-based self-interference cancellation for full-duplex broadband transmission},
  author={Wu, Jiayan and Cheng, Wenchi and Wang, Jianyu and Wang, Jingqing and Zhang, Wei},
  journal={IEEE Trans. Wireless Commun.},
  volume={23},
  number={7},
  pages={7159--7171},
  year={2024},
  publisher={IEEE}
}

@article{zhang2023ris,
  title={{RIS}-assisted self-interference mitigation for in-band full-duplex transceivers},
  author={Zhang, Wei and Wen, Ziyu and Du, Cheng and Jiang, Yi and Zhou, Bin},
  journal={IEEE Trans. Commun.},
  volume={71},
  number={9},
  pages={5444--5454},
  year={2023},
  publisher={IEEE}
}

@article{yang2023reconfigurable,
  title={Reconfigurable intelligent surface-aided full-duplex {mmWave MIMO}: {Channel} estimation, passive and hybrid beamforming},
  author={Yang, Songjie and Lyu, Wanting and Xanthos, Yunis and Zhang, Zhongpei and Assi, Chadi and Yuen, Chau},
  journal={IEEE Trans. Wireless Commun.},
  volume={23},
  number={4},
  pages={2575--2590},
  year={2024},
  publisher={IEEE}
}

@inproceedings{tewes2022full,
  title={Full-duplex meets reconfigurable surfaces: {RIS}-assisted SIC for full-duplex radios},
  author={Tewes, Simon and Heinrichs, Markus and Staat, Paul and Kronberger, Rainer and Sezgin, Aydin},
  booktitle={ICC 2022-IEEE International Conference on Communications},
  pages={1106--1111},
  year={2022},
  organization={IEEE}
}

@article{guo2023joint,
  title={Joint beamforming and power allocation for RIS aided full-duplex integrated sensing and uplink communication system},
  author={Guo, Yuan and Liu, Yang and Wu, Qingqing and Li, Xiaoyang and Shi, Qingjiang},
  journal={IEEE Trans. Wireless Commun.},
  volume={23},
  number={5},
  pages={4627--4642},
  year={2024},
  publisher={IEEE}
}

@article{le2023ris,
  title={RIS-assisted full-duplex integrated sensing and communication},
  author={Le, Quang Nhat and Nguyen, Van-Dinh and Dobre, Octavia A and Shin, Hyundong},
  journal={IEEE Wireless Commun. Lett.},
  volume={12},
  number={10},
  pages={1677--1681},
  year={2023},
  publisher={IEEE}
}

@article{wu2019towards,
  title={Towards smart and reconfigurable environment: {Intelligent} reflecting surface aided wireless network},
  author={Wu, Qingqing and Zhang, Rui},
  journal={IEEE Commun. Mag.},
  volume={58},
  number={1},
  pages={106--112},
  year={2020},
  publisher={IEEE}
}

@article{chen2023next,
  title={Next-generation full duplex networking systems empowered by reconfigurable intelligent surfaces},
  author={Chen, Yingyang and others},
  journal={IEEE Trans. Wireless Commun.},
  volume={23},
  number={6},
  pages={6045--6060},
  year={2024},
  publisher={IEEE}
}

@article{kermoal2002stochastic,
  title={A stochastic {MIMO} radio channel model with experimental validation},
  author={Kermoal, Jean-Philippe and Schumacher, Laurent and Pedersen, Klaus I and Mogensen, Preben E and Frederiksen, Frank},
  journal={IEEE J. Sel. Areas Commun.},
  volume={20},
  number={6},
  pages={1211--1226},
  year={2002},
  publisher={IEEE}
}

@article{gkizeli2014maximum,
  title={Maximum-{SNR} antenna selection among a large number of transmit antennas},
  author={Gkizeli, Maria and Karystinos, George N},
  journal={IEEE J. Sel. Topics Signal Process.},
  volume={8},
  number={5},
  pages={891--901},
  year={2014},
  publisher={IEEE}
}

@article{heath2025tri,
  title={The tri-hybrid {MIMO} architecture},
  author={Heath, Robert W and Carlson, Joseph and Deshpande, Nitish Vikas and Castellanos, Miguel Rodrigo and Akrout, Mohamed and Chae, Chan-Byoung},
  journal={IEEE Wireless Commun.},
year={2026},
  volume={33},
  number={1},
  pages={199-206},
  publisher={IEEE}
}

@article{li2026tri,
  title={Tri-hybrid beamforming for radiation-center reconfigurable antenna array: Spectral efficiency and energy efficiency},
  author={Li, Yinchen and Qi, Chenhao and Mao, Shiwen and Dobre, Octavia A},
  journal={IEEE Trans. Wireless Commun.},
   year={2026},
  volume={25},
  number={},
  pages={12263-12278},
  publisher={IEEE}
}

@article{zhao2025joint,
  title={Joint Positioning, Beamforming, and Power Allocation in Full-Duplex {MIMO} With Position-Reconfigurable Antenna Arrays},
  author={Zhao, Chengjie and Gong, Yuanzhe and Le-Ngoc, Tho},
  journal={IEEE Trans. Wireless Commun.},
  volume={25},
  pages={18750--18763},
  year={2026},
  publisher={IEEE}
}

@article{mendez2016hybrid,
  title={{Hybrid MIMO architectures for millimeter wave communications: Phase shifters or switches?}},
  author={M{\'e}ndez-Rial, Roi and Rusu, Cristian and Gonz{\'a}lez-Prelcic, Nuria and Alkhateeb, Ahmed and Heath, Robert W},
  journal={IEEE Access},
  volume={4},
  pages={247--267},
  year={2016},
  publisher={IEEE}
}

@article{gong2026subarray,
  title={Sub-array Selection Optimization for Joint Self-Interference and Multi-User Interference Suppression in {FD mMIMO}},
  author={Gong, Yuanzhe and Zhang, Yuanxing and Le-Ngoc, Tho},
  journal={arXiv:2606.12831},
  year={2026}
}

@inproceedings{gong2026JointICCE,
  title={Joint {Tx/Rx Antenna} Selection and Beamforming Optimization for Multi-User Full-Duplex {mMIMO}},
  author={Gong, Yuanzhe and Zhao, Chengjie and Le-Ngoc, Tho},
  booktitle={Accepted to appear in Proc. 11th IEEE Int. Conf.
Commun. Electron. (ICCE), Jul. 2026},
  pages={1--6},
  year={2026},
  organization={IEEE}
}

\end{document}